\documentclass[a4paper,11pt]{article}

\usepackage[T1]{fontenc}
\usepackage{authblk}
\usepackage{mathtools, amssymb, amsthm, mathrsfs}
\usepackage{stmaryrd} 
\mathtoolsset{showonlyrefs}
\usepackage{graphicx}
\graphicspath{{figures/}}
\usepackage[left=0.6667in, right=0.6667in, top=1in, bottom=1in]{geometry}
\usepackage[english]{babel}
\usepackage{csquotes}
\usepackage{tabularray}

\usepackage{titlecaps}
\usepackage[backend=biber, style=numeric-comp, sorting=nyt, sortcites, giveninits, maxbibnames=99]{biblatex}
\Addlcwords{a in of the for an or and by}
\DeclareFieldFormat{titlecase}{\titlecap{#1}}

\usepackage{xcolor,ninecolors}
\usepackage[colorlinks=true, linkcolor=blue4, citecolor=green6]{hyperref}

\newcommand*{\B}{\mathcal B}
\newcommand*{\R}{\mathbb R}
\newcommand*{\I}{\mathcal I}
\newcommand*{\Rp}{{\R_+}}
\newcommand*{\resto}[1][]{R^{#1}_\epsilon[\phi](f_\epsilon,f_\epsilon)}
\newcommand*{\dt}{\frac{d}{dt}}

\newcommand*{\phv}{\, \cdot \,} 
\newcommand*{\frza}[1][]{\frac{#1 V_a}{m + \alpha}}
\newcommand*{\frzb}[1][]{\frac{#1 V_r}{m + \beta}}

\newcommand*{\F}{\mathscr F}
\newcommand*{\BB}{\mathscr B}
\newcommand*{\CCC}{\mathscr C}
\newcommand*{\finf}{f^\infty}
\newcommand*{\generalC}[4]{\mathcal 
C_{\fontsize{7pt}{7pt}\selectfont\begin{smallmatrix} #1,\!\! & #2,\\ #3\!\! & 
#4 \end{smallmatrix}}}

\newcommand*{\CC}{\generalC{\alpha}{\lambda}{m}{}}
\renewcommand*{\(}{\begin{equation}}
\renewcommand*{\)}{\end{equation}}
\let\LE\lesssim
\DeclareMathOperator{\Prob}{Prob}

\DeclarePairedDelimiter{\ev}{\langle}{\rangle}
\DeclarePairedDelimiter{\abs}{\lvert}{\rvert}

\theoremstyle{plain}

\theoremstyle{remark}
\newtheorem{remark}{Remark}

\title{Emerging properties of the degree distribution in large non-growing networks}
\author[1]{Jonathan Franceschi\thanks{\texttt{jonathan.franceschi01@universitadipavia.it }}}
\author[1,2]{Lorenzo Pareschi\thanks{\texttt{L.Pareschi@hw.ac.uk }}}
\author[3]{Mattia Zanella\thanks{\texttt{mattia.zanella@unipv.it}}}

\affil[1]{Department of Mathematics and Computer Science, \protect\\ University of Ferrara, Ferrara, Italy}
\affil[2]{Maxwell Institute for Mathematical Sciences and Department of Mathematics, \protect\\ Heriot Watt University, Edinburgh, UK  \vspace*{2mm}}
\affil[3]{Department of Mathematics ``F. Casorati'', \protect\\ University of Pavia, Pavia, Italy \vspace*{2mm}} 

\date{}

\begin{document}
\maketitle

\begin{abstract}
The degree distribution is a key statistical indicator in network theory, often used to understand how information spreads across connected nodes. In this paper, we focus on non-growing networks formed through a rewiring algorithm and develop kinetic Boltzmann-type models to capture the emergence of degree distributions that characterize both preferential attachment networks and random networks. Under a suitable mean-field scaling, these models reduce to a Fokker--Planck-type partial differential equation with an affine diffusion coefficient, that is consistent with a well-established master equation for discrete rewiring processes. We further analyze the convergence to equilibrium for this class of Fokker--Planck equations, demonstrating how different regimes---ranging from exponential to algebraic rates---depend on network parameters. Our results provide a unified framework for modeling degree distributions in non-growing networks and offer insights into the long-time behavior of such systems.
\end{abstract}


\tableofcontents

\section{Introduction}
Many objects in the physical, biological, and social sciences may be considered as a network, representing a large collection of nodes/vertices joined together through weighted lines/edges that synthesize the connection strength between subgroups of nodes. The observation of emerging connection patterns of networks, objects that are characterized by millions of vertices, is typically obtained through the extrapolation of statistical quantities encapsulating the macroscopic properties influencing the information flow. This simplified representation, reducing an interacting system to an abstract structure, had a tremendous impact in many scientific communities in both the pure and applied fields like social and information systems, epidemic dynamics, mobility and finance. Without intending to review the very huge literature on these topics, we refer the reader to~\cite{caldarelli,colizza_etal,newman03,newman10,pastor_etal,pastor_vespignani,dutta2018bayesian,during2009boltzmann} and the references therein. The interest in networked systems gained also popularity in cognate areas where interactions in particle systems are mediated by underlying networks, see e.g.,~\cite{banda,barre_mms,barre_etal,bonnet,borsche,DFWZ} and the references therein.  

Amongst the simplest network model, the Erd\H{o}s-Rényi model describes a random graph where each node is connected with a given probability, leading to a Poisson connectivity distribution when the number of nodes becomes large. Nevertheless, many real world networks possess more complex degree distributions. In the last decades, several groundbreaking contributions shed light on  network formation processes triggering the emergence of prescribed stylized facts in terms of degree distribution and centrality indices, see e.g.,~\cite{jackson,das}. In this direction, the observation of so-called scale-free networks, i.e., networks whose degree distribution is of power-law-type, has been of paramount importance~\cite{clauset2009power}. A variety of generative algorithms have been proposed 
(see e.g.,~\cite{greenhill2021generating} and references therein)
to capture the scale-free feature of growing networks as a result of two mechanisms
\begin{itemize} 
\item[$(i)$] networks expand continuously through the addition of new nodes, 
\item[$(ii)$] the incoming nodes are connected preferentially to highly connected nodes.
\end{itemize}
We point the interested reader to~\cite{bara1,bara2,whigham2016network} for further details.
On the other hand, substantial researching effort has been devoted to non-augmentative sampling algorithms~\cite{gao2018uniform,arman2021fast}. In this case, the goal is to sample a (random) network realization starting from a given graphical degree sequence, i.e., a sequence of number of connections among the nodes. This often involves choosing nodes' pairs by assigning them a sampling likelihood based on their respective degrees and then sampling by using an acceptance/rejection procedure.

One possibility of such approaches has been given in the works
\cite{OH,Xie08} where the following rewiring algorithm has been proposed: we consider a network of $N\gg0$ nodes connected through a set of edges, and at each iteration we perform the following two steps
\begin{itemize}
\item[$(i)$] select randomly an edge connecting two nodes and remove it; 
\item[$(ii)$] add a connection preferentially between the disconnected node and a highly connected one. 
\end{itemize}
As a result of the rewiring strategy, the total number of nodes remains unchanged. We point the interested reader also to~\cite{albi_etalproc,Xie08} for more details. In the discrete case, where $c \in [1,\dots,c_{max}] \subset \mathbb N$, denoting by $\rho(c,t)$ the degree distribution of the network at time $t\ge0$, in~\cite{Albi16} the following master equation has been derived
\(\label{eq:masterequation}
\begin{aligned}
\partial_t \rho(c,t) &= \frac{2V_a}{m + \alpha}[(c- 1 +\alpha)\rho(c-1,t) - 
(c+\alpha)\rho(c,t)]\\
&{\hphantom{{}=}} + \frac{2V_r}{m }[(c+ 1)\rho(c+1,t) - 
c\rho(c,t)],
\end{aligned}
\)
where $m>0$ is the mean number of connections, $V_a,V_r>0$ are coefficients linked to the speed of attachment and removal, respectively, and $\alpha\ge0$ characterizes the preferential process. 
It is worth noting that, as described in~\cite{Xie08,Albi16}, in the case $V_a = V_r$, the unique steady state of the master equation~\eqref{eq:masterequation} is explicitly computable and reads
\(\label{eq:mastersteady}
\rho^\infty(c) = \Bigl(\frac{m}{m + \alpha}\Bigr)^c \frac{1}{c!}\, \alpha(\alpha + 1) \dots (\alpha + c - 1)\Bigl(\frac{\alpha}{\alpha + m}\Bigr)^\alpha. 
\)
Furthermore, the obtained $f^\infty(c)$ possesses two useful approximation, which are
\begin{equation}
\label{eq:approximations}
 \rho^\infty(c) \approx \frac{e^{-c}}{c!}m^c = \operatorname{Pois}(m) \quad \textrm{if $\alpha\gg 1$}, \qquad \textrm{and}\qquad \rho^\infty(c) \approx  \dfrac{\Lambda}{c}  \qquad  \textrm{if $\alpha\ll 1$},
\end{equation}
where $\Lambda$ is a normalization constant. In particular, in the case $\alpha \to 0^+$, the approximation can actually be simplified to
\(\label{eq:small-a-approx}
\rho^\infty(c) \approx \frac{\bigl(\log(c_{\mathrm{max}}) + \gamma\bigr)^{-1}}{c}, \qquad \alpha \ll 1,
\)
where $\gamma$ is the Euler--Mascheroni constant. Therefore, at the mean-field level, the proposed algorithm is consistent with a random graph in a non-preferential regime, and with a scale-free network with fixed degree distribution, in the strongly preferential case. 

In this paper, our goal is to develop an agent-based version of a large non-growing network starting from a kinetic point of view in a way that its mean-field limit is consistent with the mentioned master equation~\eqref{eq:masterequation} in suitable regimes. To this end, we formulate two distinct agent-based models (ABM) to derive from simple binary interaction rules the emerging graph topology in terms of Boltzmann-type equations. When the updates become quasi-invariant, we derive then a Fokker--Planck-type equation. This partial differential equation  expresses the evolution of the contact distribution of the network and has  computable equilibrium distribution.

Recalling classical results in kinetic theory,~\cite{bobylev_toscani,toscani99}, we establish the rate of convergence towards equilibrium. Through the study of the dissipation properties of the Shannon entropy functional, a phase transition occurs in terms of the parameter expressing the preferential  attachment dynamics. Hence, a critical parameter, depending on macroscopic properties of the network, exists and differentiates the trends to equilibrium between exponential to algebraic. In recent years a growing interest has been devoted to Fokker--Planck equation with non-constant coefficients~\cite{ATZ,Furioli17,Furioli22,Toscani21} or characterized by a  subcritical confinement~\cite{kavian,TZ21}.
 
The paper is organized as follows. In the next section, we introduce the agent-based binary dynamics and describe it using Boltzmann-type equations. We then derive its Fokker--Planck approximation, analyzing its key properties and long-term behavior. The different asymptotic states, including Poisson-type distributions and the emergence of power laws, are also discussed. We demonstrate that an alternative dynamics, which avoids the rejection mechanism, can be derived, and that its mean-field limit leads to the same Fokker--Planck equation as the original dynamics. Section~\ref{sec:trends} focuses on the long-term behavior of the limiting Fokker--Planck equation and establishes log-Sobolev-type inequalities for the convergence to equilibrium. In Section~\ref{sec:numerics}, we provide numerical evidence of the system's behavior and its ability to accurately describe the emerging degree distribution profiles in various regimes. Finally, concluding remarks are presented in the last section.

\section{Kinetic models for non-growing networks formation}
\label{sec:kinetic-models}

An increasing number of real world phenomena have been fruitfully described by kinetic and mean-field models. Particular attention has been paid in the past decade to self-organizing features of large many-agent systems, see, e.g.,~\cite{Pareschi13}. In this section, we introduce a kinetic modelling approach for modelling the formation of the contact distribution on networks composed by an infinite number of nodes. We associate to each node a number of connections, or \emph{degree}, which we denote by~$c \in \I \coloneqq\{ x \in \R \mid x > 0\}$. We are interested in studying the sequence of degrees of the network: in our setting, this equates to studying the behavior of the distribution~$f(c,t)$, such that $f(c,t)dc$ represents the fraction of nodes having degree in $[c,c + dc)$ at time $t\ge0$. 

Information on the distribution function $f(c,t)$ is then useful to obtain information about aggregate or \emph{observable} quantities. Indeed, since
\[
\int_\I f(c,t)\, dc = 1,
\]
the evaluation of
\[
\int_I \phi(c) f(c,t)\, dc
\]
gives us information about the macroscopic quantity $\phi(c)$, also known as the observable, for any smooth, compactly-supported, test functions $\phi(\phv)$. Important observable quantities are the principal moments of the density, $\phi(c) = c^n$, $n\ge1$. For instance, if $\phi(c) = c$ we obtain the evolution of the mean number of contacts for the considered (infinite) set of networks. 

The rest of the section is devoted to presenting suitable models to describe the emergence of given connection distributions. In a kinetic setting, the evolution is due to pairwise interactions, corresponding, in our case, to interactions between two nodes. In more detail, to obtain relevant  degree distributions we will present two possible strategies mimicking a classical rewiring dynamics. 

In order to leverage kinetic-type equations theory we assume 
\begin{itemize}
    \item The networks have an infinite number of nodes such that each node is undistinguishable from the others. 
    \item The number of connections $c\ge0$ is a real number, since networks are considered in a statistical setting. This allows to consider continuous distributions of nodes. 
    \item Each update to the number of connections happening via the prescribed dynamics takes place continuously, rather than in a discrete fashion. This implies that we can actually consider a time-dependent degree distribution~$f(c,t)$ which evolves continuously in both variables.
\end{itemize}

\subsection{Linear rewiring dynamics}
\label{sec:non-max}

In this first case, we suppose that at time $t > 0$ we select a pair of nodes with degrees $c$ and $c_*$. Then, they either gain a certain amount of connections or they lose it. The amount of connections is the sum of two components: one is fixed and deterministic, which we call $\delta > 0$, and one is stochastic. 

This can be expressed by the following update rules: the gain reads
\(\label{eq:update1}
\begin{aligned}
    c'   &= c + \delta + \eta_a\\
    c_*' &= c_* + \delta + \tilde\eta_a
\end{aligned}
\)
and the loss reads
\(\label{eq:update2}
\begin{aligned}
    c''   &= c - \delta + \eta_r\\
    c_*'' &= c_* - \delta + \tilde\eta_r.
\end{aligned}
\)
The variables $\eta_a$, $\tilde\eta_a$, $\eta_r$, $\tilde \eta_r$ are i.i.d.\ random variables with support $\Omega \subseteq \R$. In particular, $\eta_a$ and $\tilde \eta_a$ have probability density function $\Theta(\phv)$ such that its mean is zero and its variance is $\sigma^2$. Similarly, the density~$\Psi(\phv)$ associated to the second pair~$\eta_r$, $\tilde\eta_r$ has zero mean and variance equal to~$\sigma^2$, too. 

Given the test function $\phi \in C^\infty_c(\I)$, i.e., $\phi$ is smooth and compactly supported, the kinetic equation associated to the updates~\eqref{eq:update1}--\eqref{eq:update2} can be fruitfully written in weak form in the following fashion 
\(\label{eq:wf}
\begin{aligned}
\dt \int_\I f(c,t)\phi(c)\, dc &= \int_{\I}\int_\I \B_a(c,c_*) 
\ev{\phi(c')-\phi(c)}_{a} f(c,t)f(c_*,t) \, dc \, dc_*\\
&+ \int_\I\int_\I \B_r(c,c_*) \ev{\phi(c'')-\phi(c)}_{r} 
f(c,t)f(c_*,t)\, dc \, dc_*,
\end{aligned}
\)
where we note with $\ev{\phv}_a$ the expected value with respect to the random variable~$\eta_a$ linked to the attachment dynamics~\eqref{eq:update1}, and with  $\ev{\phv}_r$ the expected value with respect to the random variable~$\eta_r$ linked to the removal dynamics~\eqref{eq:update1}.

The operators $\B_a$ and $\B_r$ are kernels tuning the frequency of interactions between two nodes characterized by connections $c\ge0$ and $c_*\ge0$. In the following, we will consider 
\(\label{eq:non-maxwellian-kernel}
\begin{aligned}{}
    \B_a(c,c_*) &\coloneqq \Theta(\eta_a)\Theta(\tilde\eta_a) V_a\frac{(c+\alpha)(c_* +\alpha)}{(m + \alpha)^2}, \\
    \B_r(c,c_*) &\coloneqq \Psi(\eta_r)\Psi(\tilde\eta_r) \chi(c'' > 0)\chi(c_*'' > 0) V_r\frac{(c+\beta)(c_*+\beta)}{(m + \beta)^2},
\end{aligned}
\)
where $m = m(t)$ is the average degree of the network at time~$t\ge 0$, $\alpha, \beta \ge 0$ are nonnegative constants and $V_a,V_r > 0$ are constant rates as in~\cite{Albi16}. 

\begin{remark}
    The presence of a noise term in equation~\eqref{eq:update1} and especially in~\eqref{eq:update2} can lead to non-positive values for $c'$ and $c'_*$, even for bounded choices of functions $\Theta(\phv)$ and $\Psi(\phv)$. We can ensure that the post-interaction values $c', c_*'$ belong to $\I$ if we impose that $\abs{\eta_a} < \delta$ and $\abs{\tilde\eta_a} < \delta$; the same cannot be said for the removal update.  Indeed, we may observe that in \eqref{eq:update2} we cannot guarantee that $c^{\prime\prime}\ge0$ and $c_*^{\prime\prime}\ge0$ for any value of $\delta >0$, provided $c,c_*\ge0$. To this end, in \eqref{eq:non-maxwellian-kernel}, we introduced a cut-off through the indicator function $\chi(c^{\prime\prime}\ge0)$ to avoid all the interactions producing unphysical effects. 
\end{remark}

The binary updates~\eqref{eq:update1}--\eqref{eq:update2} are a generalization of the discrete algorithm proposed in~\cite{Xie08} and adapted in a kinetic-theory framework in~\cite{Albi13,Albi16}: there, a node is selected with probability weighted by~$V_a(c + \alpha)/(m(t) + \alpha)$ and, if admissible, a connection is added to it. The same procedure is performed to remove a connection, but this time the weight for the selection is given by~$V_r(c + \beta)/(m(t) + \beta)$. These characteristics make it a \emph{rewiring} algorithm, that is, the number of edges is preserved in time, along with the number of nodes. Moreover, if the relative importance of adding connections matches the one of removing them, i.e., if $V_a = V_r$, the average mean degree of the network is also preserved in time.

Our goal is to compare the properties of the generalized model~\eqref{eq:wf} with the one enjoyed by the master equation~\eqref{eq:masterequation}. In particular,
we can leverage the weak form~\eqref{eq:wf} to study the behavior of some observable quantities: for instance, plugging $\phi(c) \equiv 1$ in~\eqref{eq:wf} leads to the conservation in time of the number of nodes in the network. The choice $\phi(c) = c$, instead, would give us the evolution in time of the average degree: the presence of the nonlinear indicator functions in~\eqref{eq:non-maxwellian-kernel}, however, makes its computation unpractical, along with the analysis of moments of higher order of the solution~$f(c,t)$. In order to overcome this difficulty and more in general
in order to gain further insights on equation~\eqref{eq:wf}, like for example an explicit form for its large-time behavior, we consider in what follows its mean-field approximation obtained within the so-called quasi-invariant or \emph{grazing-collision} limit. This procedure simplifies the integro-differential Boltzmann-type model~\eqref{eq:wf} into a surrogate, Fokker--Planck-type model which should ease the computations.

\subsubsection{Mean-field approximation}

In order to derive a mean-field approximation of our model~\eqref{eq:wf}, we consider the scaling  
\[
\delta \to \epsilon\delta, \qquad \sigma \to \sqrt\epsilon\sigma,
\]
where $\epsilon>0$ is a suitable scaling coefficient. Due to the presence of 
the characteristic function in the interaction kernel, we may split our weak 
formulation as the sum of two pieces to be treated separately 
\[
\begin{aligned}
\dt \int_I f_\epsilon (c,t) \phi(c)\, dc &=
\frac{1}\epsilon \int_\I \int_\I \ev{\phi(c')-\phi(c)}\B_a(c,c_*) 
f_\epsilon(c,t)f_\epsilon(c_*,t)\, dc_*\, dc\\
&{{}+{}}
\frac{1}\epsilon \int_\I \int_\I \ev{(\phi(c'')-\phi(c))}\B_r(c,c_*) f_\epsilon(c,t)f_\epsilon(c_*,t)\, 
dc_*\, dc\\
&= \frac{1}\epsilon \int_\I \int_\I \ev{\phi(c')-\phi(c)}\B_a(c,c_*) 
f_\epsilon(c,t)f_\epsilon(c_*,t)\, dc_*\, dc\\
&{{}+{}}
\frac{1}\epsilon \int_\I \int_\I \ev{\phi(c'')-\phi(c)}\B_r(c,c_*) 
f_\epsilon(c,t)f_\epsilon(c_*,t)\, dc_*\, dc\\
&{{}+{}}
\frac{1}\epsilon \int_\I \int_\I \ev{(1-\chi(c'' > 
\delta))(\phi(c'')-\phi(c))}\B_r(c,c_*) f_\epsilon(c,t)f_\epsilon(c_*,t)\, 
dc_*\, dc\\
&= A_\epsilon[\phi](f_\epsilon, f_\epsilon) + \resto,
\end{aligned}
\]
where we noted with $f_\epsilon(c, t)$ the solution to the scaled problem and we emphasized the dependence of the removal interaction kernel on the indicator function explicitly.
Proceeding like in~\cite{Tosin21OG,Tosin21}, our aim is to show that $\resto 
\to 0$ in the limit $\epsilon \to 0^+$, so that the model is correctly 
approximated in the mean-field regime by the sole operator $A_\epsilon$, whose 
expression is investigated later on.

For the rest of the section, we take $\phi \in C_c^\infty(\I)$, following~\cite{Tosin21}. 
We start by observing that
\[
\abs{\resto} \le \frac{1}\epsilon \biggl[  \int_\I \int_\I \ev{(1-\chi(c'' 
> \delta))\abs{\phi(c'')-\phi(c)}}\B_r(c,c_*) 
f_\epsilon(c,t)f_\epsilon(c_*,t)\, dc_*\, dc \biggr]
\]
and that we can represent $\eta_r$ as $\eta_r = \sqrt\epsilon Y$, where $Y$ is 
a random variable such that $\ev Y = 0$, $\ev{Y^2} = 1$ and $\ev{\abs{Y}^3}< 
+\infty$. We may opt for an analogous choice for the variable $\tilde \eta_r$; 
from now on we will focus only on $\eta_r$, and, consequently, on $Y$ for 
convenience.

From interaction~\eqref{eq:update2}, scaled as $\delta \to \epsilon \delta$, 
we need
\[
Y < \frac{c- 2\epsilon \delta }{\sqrt \epsilon} \eqqcolon b_\epsilon(c)
\]
so that
\[
\abs Y < b_\epsilon(c) \implies c'' \in \I.
\]
If we expand in Taylor series $\phi(c'')$ about the value $c$ we obtain
\(\label{eq:abs-expansion}
\begin{aligned}
\abs{\phi(c'') - \phi(c)} &\le \abs{\phi'(c)}(\sqrt{\epsilon}\abs Y + 
\epsilon\delta)\\
                          &\hphantom{{}=}+ \abs{\phi''(c)}(\epsilon Y^2 
                          +\epsilon^2 \delta^2 + 2\sqrt \epsilon\delta \abs Y 
                          )\\
                          &\hphantom{{}=}+ \abs{\phi'''(\bar c)}(\epsilon^{3/2} 
                          \abs Y^3 + \epsilon^3 \delta^3 + 3 \epsilon^2 \delta 
                          Y^2 + 3\epsilon^{5/2}\delta^2 \abs Y ),
\end{aligned}
\)
where $\bar c\coloneqq \theta c + (1-\theta) c''$ for a suitable $\theta \in 
[0,1]$.
In estimate~\eqref{eq:abs-expansion} we can distinguish the terms which do 
not depend on $\abs Y$: the only one that is not already a~$o(\epsilon)$ is 
$\epsilon\delta$. Due to the boundedness of $\phi^{(k)}$ for $k = 0, 1, 
\ldots$, we would just need to consider the expression\footnote{We use the 
notation $a\LE b$ to indicate that there exists a positive constant $C$, 
independent on $\epsilon$, such that $a \le C\cdot b$.}
\[
\ev{(1-\chi(c'' > \delta))\epsilon \delta} = \Prob(Y \ge b_\epsilon(c)) 
\epsilon\delta \le \frac{\epsilon\delta}{b_\epsilon(c)^2} \LE Y\epsilon^2,
\]
where the first inequality follows from Chebyshev's inequality and where the 
last inequality holds for $\epsilon \le 1/2$. We get as a consequence that every term in estimate~\eqref{eq:abs-expansion} that is independent on $Y$ vanishes in the limit $\lim_{\epsilon \to 0} \abs{\resto}$.

Let us conclude by proving that the same holds for the other terms. This time, 
we leverage H\"older's inequality alongside Chebyshev's inequality to obtain 
the estimate
\[
\ev{(1-\chi(c'' > \delta))\abs{Y}^n} \le \ev{\abs{Y}^{np}}^{1/p}\Prob(Y \ge 
b_\epsilon(c))^{1/q} \LE \ev{\abs{Y}^{np}}^{1/p}\epsilon^{1/q},
\]
where $n$ is a positive integer and $p,q \in [1,+\infty)$ are 
H\"older-conjugates. Since we assume $\ev{\abs{Y}^3}$ to be bounded, H\"older's 
inequality implies that $\ev{\abs{Y}^{np}} < +\infty$ for $n = 1$, $2$ 
and~$3$ (just take $p\in [1, 3/n]$), and in particular it holds
\[
\ev{(1-\chi(c'' > \delta))\abs{Y}} \LE  \ev{\abs{Y}^{3}}^{1/3}\epsilon^{3/2} 
= o(\epsilon).
\]
These computations show that $\abs{\resto} \to 0^+$ when $\epsilon \ll 1$, as 
desired.

Hence, if we consider $\epsilon \ll 1$, we can expand in Taylor's series the difference 
$\phi(c') - \phi(c)$ as follows
\[
\phi(c') - \phi(c) = \phi'(c)(c'-c) + \frac12 \phi(c)''(c'-c)^2 + \frac16 
\phi(\tilde c)'''(c'-c)^3,
\]
where $\tilde c \in (\min\{c',c\}, \max\{c',c\})$. Then, if we replace this 
expansion in the weak form~\eqref{eq:wf} of the model, while considering the 
scaling $t \to t/\epsilon$, we get that 
\[
\begin{aligned}
\dt \int_\Rp f(c,t)\phi(c)\, dc &= \frac{V_a}{ (m(t) + \alpha)} \int_{\Rp} (c 
+ \alpha)  \Bigl[\delta\phi'(c) + \frac12 \sigma^2\phi''(c)\Bigr] f(c,t) \, dc 
\\
&+ \frac{V_r}{ (m(t) + \beta)} \int_{\Rp} (c + \beta)  \Bigl[-\delta\phi'(c) + 
\frac12 \sigma^2\phi''(c)\Bigr] f(c,t) \, dc + O(\epsilon),
\end{aligned}
\]
where the remainder term $O(\epsilon)$ comes from integrating on the whole 
$\Rp$ rather than on the interval~$\I$, and it vanishes when $\epsilon \ll 1$. 
Integrating back by parts, the equation translates into the Fokker--Planck 
equation
\(\label{eq:FP}
\begin{aligned}
\partial_t f(c,t) = \delta \partial_c\biggl[-\Bigl(\frac{V_a(c+\alpha)}{m(t) + 
\alpha} - \frac{V_r(c+\beta)}{m(t) + \beta}\Bigr)f(c,t) \biggr]
+ \frac{\sigma^2}{2}\partial^2_{c}\biggl[\Bigl(\frac{V_a(c+\alpha)}{m(t) + 
\alpha} + \frac{V_r(c+\beta)}{m(t) + \beta}\Bigr)f(c,t)\biggr],
\end{aligned}
\)
which needs to be complemented with the following no-flux boundary conditions
\(\label{eq:bound-cond}
\begin{aligned}
-\delta\Bigl(\frac{V_a(c+\alpha)}{m(t) + \alpha} - \frac{V_r(c+\beta)}{m(t) + 
\beta}\Bigr)f(c,t)
+ \frac{\sigma^2}{2}\partial_c\biggl[\Bigl(\frac{V_a(c+\alpha)}{m(t) + \alpha} 
+ \frac{V_r(c+\beta)}{m(t) + \beta}\Bigr)f(c,t) \biggr]\bigg|_{c=0}&=0,\\
\partial_c\biggl[\Bigl(\frac{V_a(c+\alpha)}{m(t) + \alpha} + 
\frac{V_r(c+\beta)}{m(t) + \beta}\Bigr)f(c,t)\biggr]\bigg|_{c=0}&=0.
\end{aligned}
\)
The obtained Fokker--Planck-type equation is the continuous version of the master equation~\eqref{eq:masterequation}. 

\subsubsection{Evolution of first and second-order moments}
\label{subsect:moments}

The surrogate model obtained as equation~\eqref{eq:FP} allows us to study the 
evolution of the first and second-order moments of its solution~$f(c,t)$, 
without the complications given by the nonlinearities due to the presence of 
characteristic functions, needed to ensure that the boundaries are not 
violated. 

First of all, integrating equation~\eqref{eq:FP} with 
respect to~$c$ we have the preservation of the total number of nodes in the 
network, that is
\[
\dt \rho(t) = \dt \int_\I f(c,t)\, dc = 0.
\]
The evolution of the average number of connections can be obtained by 
multiplying equation~\eqref{eq:FP} by~$c$ and integrate with respect to the 
variable~$c$
\(\label{eq:mean}
\begin{aligned}
\dt m(t) &= \delta\int_{\I} c\, \partial_c\biggl[-\Bigl(\frac{V_a(c+\alpha)}{m(t) + \alpha} - \frac{V_r(c+\beta)}{m(t) + \beta}\Bigr)f(c,t) \biggr]\, dc\\
&\hphantom{={}} + \frac{\sigma^2}{2}\int_{\I} 
c\,\partial^2_{c}\biggl[\Bigl(\frac{V_a(c+\alpha)}{m(t) + 
\alpha} + \frac{V_r(c+\beta)}{m(t) + \beta}\Bigr)f(c,t)\biggr]c\, dc\\
&= \delta(V_r - V_a),
\end{aligned}
\)
in view of the boundary conditions~\eqref{eq:bound-cond}. This implies that 
the average degree of the network is preserved if and only if $V_a = V_r$, 
i.e., if on average we have the same amount of connections and disconnections 
on the network. Otherwise, the mean number of connections will increase or 
decrease accordingly.

For what concerns the second order moment, instead, we can multiply 
equation~\eqref{eq:FP} by~$c^2$ and integrate with respect to~$c$ to have
\(\label{eq:energy}
\begin{aligned}
\dt E(t) &= \delta\int_{\I} c^2 
\partial_c\biggl[-\Bigl(\frac{V_a(c+\alpha)}{m(t) + \alpha} - 
\frac{V_r(c+\beta)}{m(t) + \beta}\Bigr)f(c,t) \biggr]\, dc\\
&\hphantom{{}=}
+ \frac{\sigma^2}{2}\int_{\I} 
c^2\partial^2_{c}\biggl[\Bigl(\frac{V_a(c+\alpha)}{m(t) + 
\alpha} + \frac{V_r(c+\beta)}{m(t) + \beta}\Bigr)f(c,t)\biggr]\, dc\\
&= 2\delta\Bigl[\Bigl(\frza - \frzb\Bigr) \bigl(E(t) - (\beta - \alpha) 
m(t)\bigr)\Bigr] + \sigma^2(V_a + V_r).
\end{aligned}
\)
In particular, in the case we have conservation of the average degree, and also 
in absence of diffusion, i.e., we fix $\sigma=0$, we have
\[
\dt E(t) = -\frac{V_a(\beta - \alpha)}{(m+\alpha)(m+\beta)}(E(t) - m^2),
\]
which implies the vanishing of the variance and therefore the exponential 
convergence towards a Dirac's delta centered at the average degree of the 
network.

\subsubsection{Large-time behavior}

In the case $V_a = V_r$ the mean number of connections is a conserved quantity and we can rewrite equation~\eqref{eq:FP} by introducing the linear time scale
\[
\tau = \frac{\delta }{(m + \alpha)(m + \beta)}t. 
\]
We obtain
\(\label{eq:FP-rescaled}
\begin{aligned}
\partial_\tau f(c,\tau)
&= \partial_c \biggl[-\Bigl(V_a(c+\alpha)(m + \beta) - 
V_r(c+\beta)(m + \alpha)\Bigr)f(c,\tau) \biggr]\\
&\hphantom{{}=}
+ \frac{\lambda}{2}\partial^2_{c}\biggl[\Bigl(V_a(c+\alpha)(m + \beta) + 
V_r(c+\beta)(m + \alpha)\Bigr)f(c,\tau)\biggr],
\end{aligned}
\)
where we define  $\lambda \coloneqq \sigma^2/\delta$. Following~\cite{Xie08,Albi16}, we are interested in the behavior of the 
steady state when $V_a = V_r = 1$ and $\beta = 0$. Thus, 
equation~\eqref{eq:FP-rescaled} reduces to
\(\label{eq:FP-reduced}
\partial_\tau f(c,\tau) = \partial_c \bigl[\alpha(c - m) f(c,\tau)\bigr] + 
\frac{\lambda}{2} \partial_c^2 \bigl[((2m + \alpha)c + \alpha m)f(c,\tau)\bigr],
\)
from which we can compute the large time distribution as solution of the following differential equation 
\[
\alpha(c - m) f^\infty(c) + \frac{\lambda}{2} \partial_c \bigl[((2m + \alpha)c + \alpha m)f^\infty(c)\bigr] = 0,
\]
from which we get
\begin{equation}\label{eq:steady-reduced}
\finf(c) = \CC((2m + \alpha)c + \alpha m)^{\frac{2\alpha((2m + \alpha)m + 
\alpha m)}{\lambda(2m + \alpha)^2} - 1} \exp\Bigl(-\frac{2\alpha}{\lambda(2m + \alpha)} c\Bigr),
\end{equation}
where the following constant
\[
\CC = \frac{(2\lambda\alpha)^{\frac{\alpha((2m + \alpha)m + \alpha m)}{(2m + 
\alpha)^2}}}{\Gamma\Bigl(\frac{2\alpha((2m + \alpha)m + \alpha m)}{\lambda(2m + 
\alpha)^2}, \frac{2\alpha^2 m}{\lambda(2m + \alpha)^2} \Bigr) \lambda(2(2m + 
\alpha))^{\frac{4\alpha((2m + \alpha)m + \alpha m)}{\lambda(2m + \alpha)^2} - 
1}} \exp\Bigl(-\frac{2\alpha^2 m}{\lambda(2m + 
\alpha)^2}\Bigr),
\]
is such that $\int_{\mathbb R^+}f^\infty(c)dc = 1$. 
\begin{remark}
We can observe that the equilibrium distribution~\eqref{eq:steady-reduced} is a Gamma distribution. In detail, for $\alpha\gg1 $ we get
\[
f^\infty(c) = \mathcal C_{\infty,\lambda,m}(c+m)^{\frac{4m}{\lambda} - 1} e^{-\frac{2}{\lambda} c}.
\]
On the other hand, in the regime $\alpha\ll1$
\[
f^\infty(c) \sim \mathcal C_{\alpha,\lambda,m} \dfrac{1}{c}, 
\]
exhibiting therefore power-law tails in the limit $\alpha\to 0^+$. 
\end{remark}

\subsection{Bilinear rewiring dynamics}

In this section we propose an alternate description to model~\eqref{eq:wf} that allows a random selection of interacting nodes. This corresponds to a Maxwellian-type kinetic model where the  kernel tuning the frequency of interactions is independent on $c,c_*\ge0$. We will prove that also in this case the kinetic model converges 
in the mean-field limit to the same Fokker--Planck equation~\eqref{eq:FP}.

The new approach is based on two different update schemes. Two nodes chracterized by connection $c,c_*\ge0$ are randomly selected. Hence, the post-interaction number of connections are given by one of the following binary updates. The first reads
\(\label{eq:new-update-1}
\begin{aligned}
c'   &= \biggl(1 - \frza[\delta]\biggr)c - \frac{\alpha}{m} \frza[\delta] c_* 
+ \sqrt{\frza (c + \alpha)} \eta_1,\\
c_*' &= \biggl(1 + \frac{\alpha}{m}\frza[\delta]\biggr)c_* + \frza[\delta] c + 
\sqrt{\frza (c_* + \alpha)} \tilde\eta_1,\\
\end{aligned}
\)
being $V_a,\delta, \alpha > 0$ constants defined as in Section~\ref{sec:non-max}, $m= m(t)$ the mean number of connections and $\eta_1,\tilde\eta_1$ i.i.d.\ random variables with zero mean and finite variance~$\sigma^2$; whereas the second reads
\(\label{eq:new-update-2}
\begin{aligned}
c''   &= \biggl(1 + \frzb[\delta]\biggr)c + \frac{\beta}{m} \frzb[\delta] c_* + 
\sqrt{\frzb (c + \beta)} \eta_2,\\
c_*'' &= \biggl(1 - \frac{\beta}{m}\frzb[\delta]\biggr)c_* - \frzb[\delta] c + 
\sqrt{\frzb (c_* + \beta)} \tilde\eta_2,
\end{aligned}
\)
being $V_r, \beta > 0$ constants, and $\eta_2, \tilde\eta_2$ i.i.d. random variables with zero mean and finite variance~$\sigma^2$.

To keep the notation consistent, we say that the time evolution of $f(c,t)$ 
depends on two new operators encapsulating the two processes, $Q_a^1$ and $Q_r^1$, such that in 
strong form we have
\(
\partial_t f(c,t) = Q_a^1[f,f](c,t) + Q_r^1[f,f](c,t),
\)
where 
\(
\begin{aligned}
Q_a^1[f,f](c,t) &= \int_{\Omega^2}\int_{\I} 
\Bigl(\,{}'\B_a^1\frac{1}{J_a^1}f({}'c,t)f({}'c_*,t)
                - \B_a^1 f({}'c,t)f({}'c_*,t) \Bigr)\, dc_*\, d\eta_1\, 
                d\tilde\eta_1\\
Q_r^1[f,f](c,t) &= \int_{\Omega^2}\int_{\I} \Bigl( 
\,{}''\B_r^1\frac{1}{J_r^1}f({}''c,t)f({}''c_*,t)
                - \B_r^1 f({}''c,t)f({}''c_*,t)\Bigr)\, dc_*\, d\eta_2\, 
                d\tilde\eta_2,
\end{aligned}
\)
where $({}'c,{}'c_*)$ and $({}''c,{}''c_*)$ are the pre-collisional numbers of connections given by the pair $(c,c_*)$ after the update. Moreover, we indicate with $J_a^1$ and $J_r^1$ the Jacobian matrices associated to the transformation $({}'c,{}'c_*) \to (c,c_*)$ following interaction rules~\eqref{eq:new-update-1} 
and~\eqref{eq:new-update-2}. Finally, we consider the kernels $\B_a^1$ and $\B_r^1$ 
to be of the form
\(
\begin{aligned}
\B_a^1 &= \Theta(\eta_1)\Theta(\tilde\eta_1) \chi(c' > \delta) \chi(c_*' > \delta),\\
\B_r^1 &= \Xi(\eta_2)\Xi(\tilde\eta_2) \chi(c'' > \delta) \chi(c_*'' > \delta),
\end{aligned}
\)
where $\Theta$ and $\Xi$ are symmetric probability densities with zero mean 
and variance $\sigma^2$.

If we consider the weak form of the model arising from the microscopic update 
rules~\eqref{eq:new-update-1}--\eqref{eq:new-update-2} we obtain
\(\label{eq:new-wf}
\begin{aligned}
\dt \int_\I f(c,t)\phi(c)\, dc &=  \frac12\int_\I\int_\I\B_a^1 \bigl(\ev{\phi(c') + \phi(c_*')
    - \phi(c) - \phi(c_*)} \bigr) f(c,t)f(c_*,t) \, dc \, dc_*\\
                                &+ \frac12 \int_\I\int_\I \B_r^1 
                                \bigl(\ev{\phi(c'') + \phi(c_*'')
    - \phi(c) - \phi(c_*)} \bigr) f(c,t)f(c_*,t)\, dc \, dc_*.
\end{aligned}
\)
By replacing $\phi(c)\equiv 1$ in equation~\eqref{eq:new-wf} like done in 
Section~\ref{sec:non-max},  we can deduce again that the total number of connections is 
preserved in time. However, we face the same problem as before: we cannot easily study the evolution of the average degree of the network (or of higher order moments of $f(c,t)$) because of the nonlinear indicator functions 
that intervene to prevent boundary violations.

If we perform the same computation steps of Section~\ref{sec:non-max} on 
equation~\eqref{eq:new-wf} and consider the following scaled quantities
\[
\delta \to \epsilon\delta, \qquad \sigma \to \sqrt{\epsilon}\sigma,
\]
we can rewrite the weak form~\eqref{eq:new-wf} as the sum of two pieces
\[
\begin{aligned}
\dt\int_{\I} f_\epsilon(c,t)\phi(c)\, dc
&= \frac1\epsilon \int_\I\int_\I \B_a^1(c,c_*) \ev{\phi(c')-\phi(c)}
    f_\epsilon(c,t)f_\epsilon(c_*,t)\, dc\, dc_*\\
&+ \frac1\epsilon \int_\I\int_\I \B_r^1(c,c_*) \ev{\phi(c'')-\phi(c)}
    f_\epsilon(c,t)f_\epsilon(c_*,t)\, dc\, dc_*\\
&= \frac1\epsilon \int_\I\int_\I \B_a^1(c,c_*)\ev{\phi(c')-\phi(c)}
    f_\epsilon(c,t)f_\epsilon(c_*,t)\, dc\, dc_*\\
&+ \underbrace{\frac1\epsilon \int_\I\int_\I \B_a^1(c,c_*)\ev{(1-\chi(c'> \delta)) \phi(c')-\phi(c)} f_\epsilon(c,t)f_\epsilon(c_*,t)\, dc\, dc_*}_{\resto[a]}\\
&+ \frac1\epsilon \int_\I\int_\I \B_r^1(c,c_*)\ev{\phi(c'')-\phi(c)}
    f_\epsilon(c,t)f_\epsilon(c_*,t)\, dc\, dc_*\\
&+ \underbrace{\frac1\epsilon \int_\I\int_\I \B_r^1(c,c_*)\ev{(1-\chi(c''> \delta))\phi(c'')-\phi(c)} f_\epsilon(c,t)f_\epsilon(c_*,t)\, dc\, dc_*}_{\resto[r]}\\
&= A^1_\epsilon[\phi](f_\epsilon, f_\epsilon) + \resto[a]  + \resto[r],
\end{aligned}
\]
where again we highlight the explicit dependence of the kernels on the characteristic function. We proceed in the same spirit of Section~\ref{sec:non-max} to show that $\resto[a]$ and $\resto[r]$ vanish in the limit $\epsilon \to 0^+$; finally, we investigate the exact form of remaining operator.

In this case, in a fashion similar to our approach on the operator $Q_1$ in 
Section~\ref{sec:non-max}, we can give sufficient conditions on the random 
variable $\eta_2$ such that $\resto[r] \to 0$ whenever $\epsilon \to 0^+$. 
Indeed, if we require $\abs{\eta_2} < \delta\sqrt{ V_r}$ and $\abs{\tilde \eta_2} < \delta\sqrt{ V_r}$ we have
\[
\delta^3 V_r + \beta\delta^2 V_r + \delta\sqrt{V_r}(\delta + \beta)\eta_2 > 0,
\]
which (recalling that $m\ge \delta$ by definition of average) implies
\[
c \Bigl(1 + \frzb[\delta]\Bigr) + \frac\beta m \frzb[\delta] c_* + \sqrt{\frzb 
(c+ \beta)}\eta_2 > \delta,
\]
i.e., using update~\eqref{eq:new-update-1},  $c' > \delta$,  or, equivalently, $\resto[r] \to 0$ whenever $\epsilon \ll 1$.

Next, we investigate the term $\resto[a]$. This time we follow the same 
approach of the previous section and represent $\eta_1$ as $\eta_1 = 
\sqrt\epsilon Y_1$, where again $Y_1$ is a random variable such that $\ev {Y_1} = 0$, $\ev{Y_1^2} = 1$ and $\ev{Y_1^3} < + \infty$. From 
update~\eqref{eq:new-update-2} we know that to have an admissible interaction 
we need
\[
\abs {Y_1} \le \frac{\bigl( 1 - \frza[\epsilon\delta] \bigr)c - \frac\alpha m 
\frza[\epsilon\delta] c_*}{\epsilon\sqrt{\frza (c + \alpha)}} \eqqcolon \bar 
b_\epsilon(c)
\]
Expanding in Taylor series $\phi(c'')$ about the value $c$ we have
\(\label{eq:new-abs-expansion}
\begin{aligned}
\abs{\phi(c'') - \phi(c) } &\LE \abs{\phi'(c)}\biggl(\frza[\epsilon\delta] 
\Bigl(c + \frac\alpha m c_* \Bigr) + \sqrt{\frza (c+\alpha)}\, 
\abs{Y_1}\biggr)\\
&\phantom{{}=}+ 
\abs{\phi''(c)}\left(\Bigl(\frza[\epsilon\delta]\Bigr)^2\Bigl(c+ \frac\alpha m 
c_*\Bigr)^2 + \epsilon\frza(c+\alpha)\abs{Y_1}^2\right.\\
&\phantom{{}= \abs{\phi''(c)}}+ \left. 
2\delta^2\Bigl(\frza[\epsilon]\Bigr)^{3/2}(c + \frac\alpha m 
c_*)(c+\alpha)\abs{Y_1}\right)\\
&\phantom{{}=}+ 
\abs{\phi'''(c)}\left(\Bigl(\frza[\epsilon\delta]\Bigr)^3\Bigl(c+ \frac \alpha 
m c_*\Bigr)^3 + \epsilon^3\Bigl(\frza(c+\alpha)\Bigr)^{3/2} \abs{Y_1}^3\right.\\
&\phantom{{}= \abs{\phi'''(c)}}+ \left. 
3\delta^2\Bigl(\frza[\epsilon]\Bigr)^{3/2}\Bigl(c+\frac\alpha m 
c_*\Bigr)^2\sqrt{(c+\alpha)}\abs{Y_1}\right.\\
&\phantom{{}= \abs{\phi'''(c)}}+ 
\left.3\delta\Bigl(\frza[\epsilon]\Bigr)^{2}\Bigl(c+\frac\alpha m c_*\Bigr) 
(c+\alpha)\abs{Y_1}^2\right).\\
\end{aligned}
\)
Now we proceed like we did in Section~\ref{sec:non-max} and notice that the 
only term in estimate~\eqref{eq:new-abs-expansion} that is both independent 
from $\abs{Y_1}$ and is not already a $o(\epsilon)$ 
is~$\frza[\epsilon\delta]\Bigl(c+ \frac\alpha m c_*\Bigr)$ which can be 
regarded as satisfying
\[
\frza[\epsilon\delta]\Bigl(c+ \frac\alpha m c_*\Bigr)\LE \frza[\epsilon\delta]
\]
since we can restrict ourselves in considering compactly supported test 
functions~$\phi(\phv)$ like we did in the previous case. Then, we can apply 
once more Chebyshev's inequality to obtain
\[
\ev*{(1 - \chi(c'' > \delta ))\frza[\epsilon\delta]} = \Prob(Y_1 \ge \bar 
b_\epsilon(c))\frza[\epsilon\delta] \le \frza[\epsilon\delta]\frac{1}{\bar 
b_\epsilon(c)^2} \LE \abs{Y_1} \epsilon^2.
\]
This implies that we are left with the terms in 
estimate~\eqref{eq:new-abs-expansion} that depend on~$\abs{Y_1}$. In the 
same fashion of Section~\ref{sec:non-max}, we may choose $p\in [1,3/n]$ and 
apply again Chebyshev's and H\"older inequalities on the 
term~$\ev{\abs{Y_1}^{np}}$ for $n = 1, 2, 3$ to obtain
\[
\ev{(1-\chi(c''> \delta ))\abs{Y_1}} \LE \ev{\abs{Y_1}^3}^{1/3}\epsilon^{3/2} = 
o(\epsilon),
\]
since we are supposing that $\abs{\ev{Y_1^3}}$ is bounded.

This way we proved that every term in estimate~\eqref{eq:new-abs-expansion} 
is a~$o(\epsilon)$ and therefore $\abs{\resto[a]} \to 0$ in the limit 
$\epsilon \to 0^+$, as desired.

We proceed as in the previous section and take $\epsilon \ll 1$, so we can 
expand in Taylor's series the difference $\phi(c') - \phi(c)$ as
\[
\phi(c') - \phi(c) = \phi'(c)(c'-c) + \frac12 \phi''(c)(c'-c)^2 + \frac16 
\phi'''(\tilde c)(c'-c)^3,
\]
where $\tilde c \in (\min\{c',c\}, \max\{c',c\})$. Then, if we replace this 
expansion in the weak form~\eqref{eq:new-wf} of the model, while considering 
the scaling $t \to t/\epsilon$, we get
\[
\begin{aligned}
\dt \int_\Rp f(c,t)\phi(c)\, dc &= \int_{\Rp} 
\Bigl[-\frza[\delta]\phi'(c)\Bigl(c + \frac\alpha m\Bigr) 
+\frac{\sigma^2}{2}\phi''(c)\Bigl(\frac{V_a(c+\alpha)}{m+\alpha}\Bigr)\Bigr)f(c_*,t)
 f(c,t) \, dc_*\, dc \\
&\hphantom{={}} \int_{\Rp} \Bigl[\frzb[\delta]\phi'(c)\Bigl(c + \frac\beta 
m\Bigr) 
+\frac{\sigma^2}{2}\phi''(c)\Bigl(\frac{V_r(c+\beta)}{m+\beta}\Bigr)\Bigr)f(c_*,t)
 f(c,t) \, dc_*\, dc
\end{aligned}
\]
Integrating back by parts, the equation translates precisely into the 
Fokker--Planck equation~\eqref{eq:FP}
\[
\partial_t f(c,t) = \delta \partial_c\biggl[-\Bigl(\frac{V_a(c+\alpha)}{m(t) + 
\alpha} - \frac{V_r(c+\beta)}{m(t) + \beta}\Bigr)f(c,t) \biggr]
+ \frac{\sigma^2}{2}\partial^2_{c}\biggl[\Bigl(\frac{V_a(c+\alpha)}{m(t) + 
\alpha} + \frac{V_r(c+\beta)}{m(t) + \beta}\Bigr)f(c,t)\biggr],
\]
which needs again to be complemented with the same no-flux boundary 
conditions~\eqref{eq:bound-cond}
\[
\begin{aligned}
-\delta\Bigl(\frac{V_a(c+\alpha)}{m(t) + \alpha} - \frac{V_r(c+\beta)}{m(t) + 
\beta}\Bigr)f(c,t)
+ \frac{\sigma^2}{2}\partial_c\biggl[\Bigl(\frac{V_a(c+\alpha)}{m(t) + \alpha} 
+ \frac{V_r(c+\beta)}{m(t) + \beta}\Bigr)f(c,t) \biggr]\bigg|_{c=0}&=0,\\
\partial_c\biggl[\Bigl(\frac{V_a(c+\alpha)}{m(t) + \alpha} + 
\frac{V_r(c+\beta)}{m(t) + \beta}\Bigr)f(c,t)\biggr]\bigg|_{c=0}&=0.
\end{aligned}
\]

Now that we have shown that both a linear kinetic rewiring process and a bilinear model originate the same mean-field limit in the form of the Fokker--Planck equation~\eqref{eq:FP}. Hence, in the mean field limit, the equation of the main moments in Section \ref{subsect:moments} is consistent in both the considered modelling approaches. 

\subsection{Consistency with the master equation of the rewiring algorithm}
Let us consider the following one-dimensional Fokker--Planck equation
\(\label{eq:tobediscretized}
\partial_t u(c,t) = \partial_c [A(c,t) u(c,t)] + \partial^2_c [D(c,t)u(c,t)]
\)
with general drift and diffusion coefficients $A(c, t)$ and $D(c,t)$
and let us discretize it using grid steps $\Delta c, \Delta t > 0$ using the 
schemes
\[
\begin{aligned}
    \partial_t u &= \frac{u_i^{n+1} - u_i^n}{ \Delta t},\\
    \partial_c [Au] &= \frac{A_{i+1}^n u_{i+1}^n - A_{i-1}^n u_{i-1}^n}{2\Delta 
    c},\\
    \partial_c^2 [Du] &= \frac{D_{i+1}^n u_{i+1}^n - 2 D_i^n u_i^n + D_{i-1}^n 
    u_{i-1}^n}{(\Delta c)^2}.
\end{aligned}
\]
We obtain
\(\label{eq:general-discrete}
\frac{u_i^{n+1} - u_i^n}{ \Delta t} = \frac{A_{i+1}^n u_{i+1}^n - A_{i-1}^n 
u_{i-1}^n}{2\Delta c} + \frac{D_{i+1}^n u_{i+1}^n - 2 D_i^n u_i^n + D_{i-1}^n 
u_{i-1}^n}{(\Delta c)^2},
\)
where $u_i^n = u(c_0 + i\Delta c, t_0 + n\Delta t )$. We have that 
equation~\eqref{eq:tobediscretized} corresponds to equation~\eqref{eq:FP} 
with
\[
\begin{aligned}
u(c,t) &= f(c,t)\\
A(c,t) &= -\delta \Bigl(\frac{V_a(c+\alpha)}{m(t) + \alpha} - 
\frac{V_r(c+\beta)}{m(t) + \beta}\Bigr)\\
D(c,t) &= \frac{\sigma^2}2\Bigl(\frac{V_a(c+\alpha)}{m(t) + \alpha} + 
\frac{V_r(c+\beta)}{m(t) + \beta}\Bigr).
\end{aligned}
\]
Replacing those terms in equation~\eqref{eq:general-discrete} and choosing 
$\Delta c = \delta = \sigma = 1$ and with $c \in \{1, 2, \ldots, c_{\mathrm{max}}\}$,  we obtain exactly the master equation~\eqref{eq:masterequation}, as desired.

\begin{figure}[htbp]
    \centering
    \hbox to \textwidth{%
    \includegraphics[width=0.45\textwidth]{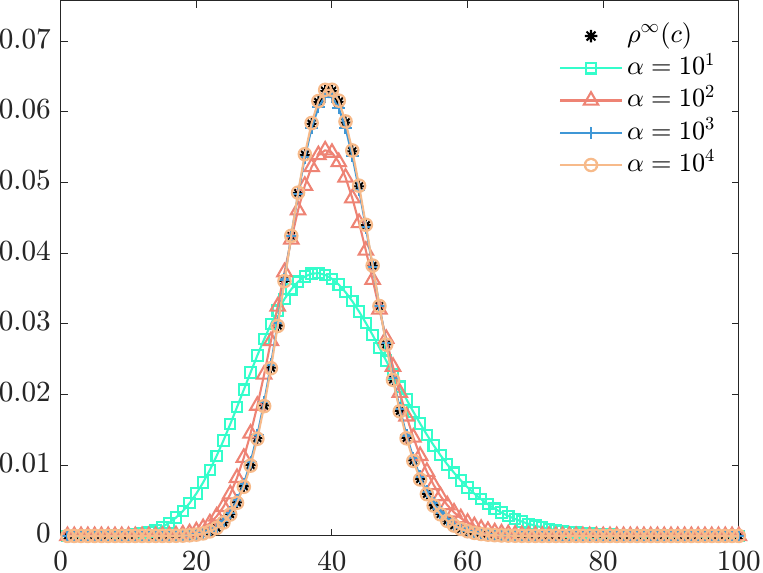}\hfil
    \includegraphics[width=0.45\textwidth]{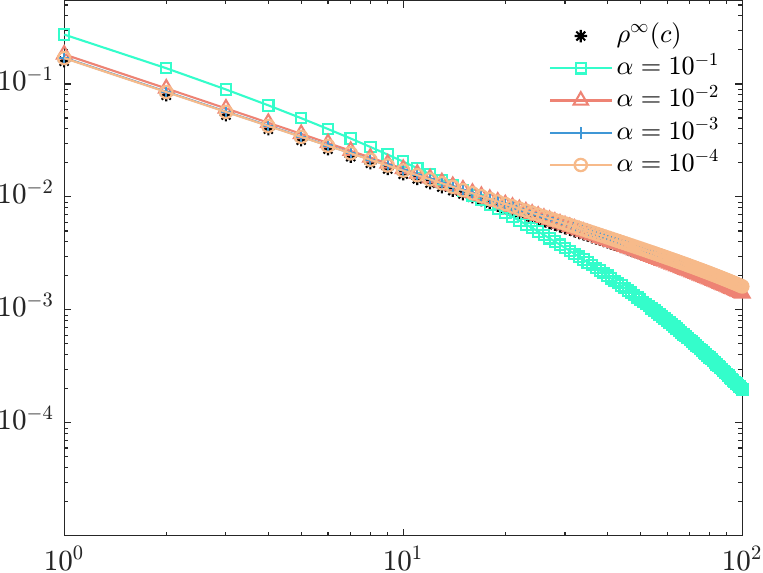}}
    \caption{Comparison of the large time-behavior of the discretization~\eqref{eq:general-discrete} of equation~\eqref{eq:tobediscretized} with the approximations given by~\eqref{eq:approximations} and~\eqref{eq:small-a-approx}. In both cases we fixed $V_a = V_r = 1$, $\beta=0$, $m = 40$ and $c_{\mathrm{max}} = 300$. On the left, convergence toward the Poisson distribution $\operatorname{Pois}(m)$ for large values of~$\alpha$. On the right, convergence toward the power-law distribution~\eqref{eq:small-a-approx} for small values of~$\alpha$.}
    \label{fig:discrete}
\end{figure}

To further support this derivation, we report in Figure~\ref{fig:discrete} the results of simulating the discretized equation~\eqref{eq:general-discrete}: it shows the convergence of the large-time behavior of the numerical solution to the steady states computed in~\eqref{eq:approximations}--\eqref{eq:small-a-approx} for large and small values of the parameter~$\alpha$. We considered as initial datum a Dirac's Delta function centered at the initial average degree $m = 30$, with discretization steps $\Delta c = 1$ and $\Delta t = 10^{-4}$. 

Equation~\eqref{eq:wf} is not the only one that can generate the mean-field limit~\eqref{eq:FP}. Model~\eqref{eq:wf} is of non-Maxwellian type, i.e., the frequency of interactions between nodes is weighted by a function of their degrees, while the update is constant. On the other hand, we can consider a model in which the frequency of interactions is constant, while the update consists of a (nonlinear) function of their degrees. This is known as a Maxwellian model.

\section{Trends to equilibrium for the Fokker--Planck rewiring equation}
\label{sec:trends}

The obtain mean-field equation~\eqref{eq:FP} may be rewritten in a simplified form as follows
\(\label{eq:affine-FP}
\partial_t f(c,t) = A\partial_c[(c - \mu)f(c,t)] + \partial_c^2[(Bc + C)f(c,t)],
\)
where
\(\label{eq:substitutions}\renewcommand{\arraystretch}{2}
\begin{array}{l@{\extracolsep{1em}}l}
A = V_a(m + \beta) - V_r(m + \alpha), &
B = V_a(m + \beta) + V_r(m + \alpha),\\
C = \alpha V_a(m + \beta) + \beta V_r(m + \alpha), &
\,\mu = -\frac{\alpha V_a(m + \beta) + \beta V_r(m + 
\alpha)}{V_a(m + \beta) - V_r(m + \alpha)}. 
\end{array}
\)
coupled to the following no-flux boundary conditions
\(\label{eq:affine-bc}
\begin{aligned}
A(c - \mu)f(c,t) + \partial_c[(Bc + C)f(c,t)]\bigg|_{c=0}&=0,\\
\partial_c[(Bc + C)f(c,t)]\bigg|_{c=0}&=0
\end{aligned}
\)

Next, we investigate the trend to equilibrium of solutions of~\eqref{eq:affine-FP}. In doing so, we are interested in the evolution in time of its Shannon entropy relative to the steady-state density $f^\infty(c)$ defined in~\eqref{eq:steady-reduced}. Following~\cite{Furioli17}, given two densities $f(c,t)$, $g(c)$, $c \in \mathbb R_+$ we can define the Shannon entropy as follows
\(\label{eq:def-entropy}
H(f \mid f^\infty)(t) = \int_{\mathbb R_+} f(c,t) \log \frac{f(c,t)}{f^\infty(c)}\, dc.
\)
We observe that $H(f\mid f^\infty)\ge0$ and $H(f \mid f^\infty) = 0$ if and only if  $f(x) = \finf(x)$. Hence, to prove the convergence towards the equilibrium density we can typically observe that, if $f(c,t)$ solution to~\eqref{eq:affine-FP} is sufficiently regular
\begin{equation}
\label{eq:Hevo}
\dfrac{d}{dt}H(f\mid f^\infty)(t) = \int_{\mathbb R_+} \left(1+\log \dfrac{f(c,t)}{f^\infty(c)} \right)\partial_t f(c,t)dc = -I_H(f\mid f^\infty)(t), 
\end{equation}
being $I_H(\cdot\mid\cdot)$ the Fisher information. Following Theorem 7 in~\cite{Furioli17} we can further prove that the entropy production term is given by 
\[
I_H(f\mid f^\infty)(t) = 4 \int_{\mathbb R_+} (Bc + C)f^\infty(c) \left( \partial_c \sqrt{\dfrac{f(c,t)}{f^\infty(c)}}\right)^2 dc, 
\]
which guarantees a monotone decay of the entropy. 

Next, if we define the Hellinger distance between two densities $f(c,t)$, $f^\infty(c)$ as follows
\(\label{eq:def-Hellinger}
d_H^2(f, f^\infty) = \int_{\Rp} \left(\sqrt{f(c,t)} - \sqrt{f^\infty(c)}\right)^2\, dc. 
\)
If $f(c,t)\ge0$ is solution to~\eqref{eq:affine-FP}, the evolution  of the Hellinger distance can be obtained from Theorem 7 of~\cite{Furioli17} and reads
\begin{equation}
\label{eq:d_diss}
\dfrac{d}{dt}d^2_H(f\mid f^\infty) =  - 8 \int_{\mathbb R_+}(Bc + C)f^\infty(c) \left( \partial_v\sqrt[4]{\dfrac{f(c,t)}{f^\infty(c)}} \right)^2 \le0, 
\end{equation}
which provides monotone decay of $d_H^2$.  From the Cauchy-Schwarz inequality we can also bound the introduced distance as follows
\begin{equation}
\label{eq:dH_ineq}
    d_H^2(f \mid f^\infty) \le 2\Bigl(1 - \Bigl(\int_{\Rp}\sqrt{f(c,t)f^\infty(c)}\, dc \Bigr)^2\Bigr).
\end{equation}
Following~\cite{Furioli17} we can prove that a weighted Poincaré inequality  is available, from which we get
\begin{equation}\label{eq:IH_ineq}
    I_H(f\mid \finf) \ge 4\Bigl(1 - \Bigl(\int_{\Rp}\sqrt{f(c,t)\finf(c)}\, dc\Bigr)^2\Bigr),
\end{equation}
Hence, coupling~\eqref{eq:dH_ineq} with~\eqref{eq:IH_ineq} we get
\(\label{eq:Fisher-Hellinger}
I_H(f\mid \finf) \ge 2 d_H^2(f \mid \finf).
\)
Inequality~\eqref{eq:Fisher-Hellinger} coupled with ~\eqref{eq:Hevo} allows to conclude that
\[
    \frac{d}{dt} H(f\mid \finf) \le -2d_H^2(f\mid \finf).
\]
Hence, after time integration we get
\[
2 \int_0^t d_H^2(s)ds < H(f\mid f^\infty)(0),
\]
and therefore $d_H^2 \in L^1([0,+\infty)]$ and from~\eqref{eq:d_diss} is monotonically decreasing. Finally, we get that $d^2_H(f\mid f^\infty)$ is $o(t^{-1})$ which provides at least algebraic decay towards the steady state.
It is worth to notice that the convergence of $d_H$ implies the convergence in $L^1(\mathbb R_+)$ since from the Cauchy-Schwartz inequality we have
\[
\int_{\mathbb R_+}|f(c,t)-f^\infty(c)|dc \le 2d_H(f\mid f^\infty).
\]
We can actually achieve far better convergence rate, provided a log-Sobolev inequality is available for the Fokker--Planck equation of interest. 

\subsection{Log--Sobolev inequalities for Fokker--Planck rewiring equation}

The idea is to follow~\cite{Toscani21,Furioli17,BakryEmery} and leverage log-Sobolev inequalities arising from Fokker--Planck type equations with general drift coefficient but constant diffusion. Indeed, it is known (\cite{Otto00}) that the Shannon entropy of the solution relative to the steady state can be bounded by the unitary Fisher information, provided that the potential associated to the drift coefficient of the Fokker--Planck type equation is uniformly convex. In other words, if we have the following equation
\(\label{eq:constant-diffusion-FP}
\partial_t h(x,t) = \partial_{x}^2 h(x,t) + \partial_x (a'(x)h(x,t)), \qquad x \in \Rp,
\)
and such that the potential $a(x)$ satisfy
\[
h^\infty(x) = C e^{-a(x)},
\]
for a suitable normalizing constant $C> 0$, and at the same time we have
\[
a''(x) \ge \rho > 0, 
\]
then the following log-Sobolev-type inequality holds
\[
    H(h\mid h^\infty) \le \frac{1}{2\rho} I_H(h\mid h^\infty).
\]
Such bound, coupled with inequality~\eqref{eq:Hevo}, allows to obtain a Gronwall-type inequality for the derivative of the Shannon relative entropy of the solution,  proving thus exponential convergence to equilibrium density. Therefore, our goal is to work with a Fokker--Planck type equation with constant diffusion related to our affine equation~\eqref{eq:affine-FP} such that they are of equivalent form and such that the resulting potential for the drift term of this new equation is also uniformly convex.

We start by introducing the adjoint equation to~\eqref{eq:affine-FP}: in particular, we 
consider the function~$F(x,t) \coloneqq f(c,t)/\finf(c)$ which is solution to
\(\label{eq:adjoint}
\partial_t F(c,t) = -A(c - \mu) \partial_c F(c,t) + (Bc + C)\partial_c^2 
F(c,t). 
\)
Now we introduce a change of variable, see~\cite{Feller52,Furioli17}, with the aim to obtain an equation with constant diffusion.  
To this end, we may consider the following change of variables
\(\label{eq:change}
y(c) \coloneqq \frac{2\sqrt{Bc + C}}{B}, \qquad c(y) = \frac{B^2y^2 - 4C}{4B}, 
\qquad B > 0,
\)
and its associate distribution $G(y,t)$, which we impose to satisfy $G(y,t) = 
F(x,t)$. Hence, we compute 
\(\label{eq:derivatives}
\begin{aligned}
    \partial_c F(c,t) &= \frac{\partial G(y,t)}{\partial y} \frac{dy(c)}{dc} = 
    \partial_y G(y,t) \frac{1}{\sqrt{Bc + C}},\\
    \partial_c^2 F(c,t) &= \partial_c (\partial_c F(c,t)) = \partial_y^2 
    G(y,t)\frac{1}{Bc + C} - \partial_y G(y,t) \frac{B}{2(Bc + C)^{3/2}},
\end{aligned}
\)
Plugging the obtained terms into equation~\eqref{eq:adjoint} we 
obtain
\[
\partial_t G(y,t) = -\frac{2A(c - \mu) + B}{2\sqrt{Bc + C}}\partial_y G(y,t) + 
\partial_y^2 G(y,t),
\]
or equivalently
\(\label{eq:g-expanded}
\partial_t G(y,t) = -\frac{2A(B^2y^2 - 4C - 4B\mu) + 4B^2}{4B^2 y}\partial_y 
G(y,t) + \partial_y^2 G(y,t).
\)
Equation~\eqref{eq:g-expanded} can then be rewritten in the form
\(\label{eq:g-potential}
\partial_t G(y,t) = \partial_y \bigl( -W'G(y,t) +\partial_y G(y,t)\bigr),
\)
where $W'$ is the associated potential
\(\label{eq:W'}
W' \coloneqq \frac{A}{2}y - \frac{2AC + B(2A\mu - B)}{B^2 y}.
\)
\begin{remark}
The change of variable~\eqref{eq:change} implies that $y(c)$ belongs to the 
ray $[2\sqrt{C}/B, +\infty)$, so that the steady state of 
equation~\eqref{eq:g-potential} $g_\infty(y) = K\exp(-W(y))$, up to a 
suitable multiplicative positive constant~$K$, is defined on the complete 
manifold~$[2\sqrt{C}/B, +\infty)$.
\end{remark}
Now, in order to leverage the argument of Bakry and Emery~\cite{BakryEmery} 
we need to prove that the potential~$W$ is strictly convex. If we differentiate again the function~$W$ with respect to~$y$ we have
\(\label{eq:W''}
W'' = \frac{A}{2} + \frac{2AC + B(2A\mu - B)}{B^2}\frac{1}{y^2}.
\)
The minimum of $W''$ depends on the fraction on the right of~\eqref{eq:W''}, 
which in turn depends on the quantity $2\sqrt  C/B$, that is the minimum value 
attainable by~$y$. If we study the sign of the numerator on the right 
in~\eqref{eq:W''}, in order for it to be non-negative we obtain the following 
condition on~$A$
\(\label{eq:convexity-condition}
A \ge \frac{1}{2}\frac{B^2}{B\mu + C}.
\)
If the condition~\eqref{eq:convexity-condition} holds, then the 
potential~$W(y)$ is strictly convex with constant $\kappa = A/2$, attained 
in the limit for vanishing values of~$y$.

Recalling the parameters in~\eqref{eq:substitutions}, the 
potential is strictly convex if $\alpha$ is such that
\(\label{eq:uniform-alpha}
\alpha \ge \bar\alpha \coloneqq \frac{2m(\lambda - 2m + 2\sqrt{\lambda m + m^2}\,)}{8m - \lambda} > 
0,
\)
since we are considering that $m > \lambda$. Therefore, in all the regimes where~\eqref{eq:uniform-alpha} holds, the following log-Sobolev inequality holds
\(\label{eq:log-Sobolev-alpha}
H(f\mid \finf) \le \frac{4}{\alpha} I_H(f\mid \finf),
\)
from which we get
\[
\frac{d}{dt} H(f\mid \finf) \le - \frac{\alpha}{4} H(f\mid \finf).
\]
and from the Gronwall inequality 
\[
H(f\mid f^\infty) \le e^{-\frac{\alpha}{4} t}H(f\mid f^\infty)(0)
\]
The previous considerations allow us to conclude that the solution to our mean-field model~\eqref{eq:FP} converges exponentially to its equilibrium density provided that~\eqref{eq:uniform-alpha} holds.

\section{Numerical results}\label{sec:numerics}

In this section we present several numerical tests to complement the theoretical findings we described in the previous parts of the manuscript. We start by showing that the both the Maxwellian and the non-Maxwellian, Boltzmann-type models~\eqref{eq:wf}--\eqref{eq:new-wf} are consistent with the mean-field limit~\eqref{eq:FP} for values of the scaling parameter~$\epsilon$ small enough and for various values of the parameter~$\alpha$. This is done resorting to the direct simulation Monte-Carlo method (DSMC) where we leverage the Nanbu-Babovsky algorithm to simulate both the model with and without interaction kernel and comparing the numerical solutions obtained for large times with the analytical steady states~\eqref{eq:steady-reduced}. Next, we look at the different transients originated by the Maxwellian and the non-Maxwellian case, i.e., we show the evolution in time of the numerical distributions for various values for~$\epsilon$ and~$\alpha$. Finally, we compute the evolution in time of the Shannon entropy of the numerical solution relative to the analytical steady state, in order to investigate its decay for the same values of the parameter $\alpha$, all consistent with the bound~\eqref{eq:uniform-alpha} which implies  exponential convergence toward equilibrium. This is done within a deterministic setting, leveraging an asymptotic-preserving numerical scheme capable of simulate the density~$f(c,t)$ with arbitrary precision. The results are then compared among them to highlight the role of the parameter $\alpha$ in the decay rate.

\subsection*{Test 1: Consistency of the mean-field limit}

For this this we leverage direct simulation Monte Carlo methods for the 
Boltzmann equation; we refer to~\cite{Pareschi13,Pareschi19} and references therein for further details. In particular, we focus on model~\eqref{eq:wf}, whose implementation is more challenging due to the presence of a non-constant interaction kernel.  
We start by recalling the strong form associated to~\eqref{eq:wf}:
the overall dynamics can be written as an integro-differential Boltzmann-type equation having the form
\[
\partial_t f = Q_a(f,f) + Q_r(f,f),
\]
where $Q_a$, $Q_r$ are collisional operators encapsulating the binary updates~\eqref{eq:update1}--\eqref{eq:update2}, so that we can write
\(\label{eq:strong}
\begin{aligned}
\partial_t f(c,t) = {}& \int_{\Omega^2}\int_{\I} 
\Bigl(\,{}'\B_a\frac{1}{J_a}f({}'c,t)f({}'c_*,t)
                   -\B_a f({}'c,t)f({}'c_*,t) \Bigr)\, dc_*\,\eta_a\,\tilde\eta_a\\
                 {}+{}& \int_{\Omega^2}\int_{\I} \Bigl( 
                 \,{}''\B_r\frac{1}{J_r}f({}''c,t)f({}''c_*,t)
                   -\B_r f({}''c,t)f({}''c_*,t)\Bigr)\, dc_*\, \eta_r\, \tilde\eta_r,
\end{aligned}
\)
where $({}'c,{}'c_*)$ and $({}''c,{}''c_*)$ are the pre-collisional numbers of connections given by 
the pair $(c,c_*)$ after the update. Moreover, we indicate with $J_a$ and $J_r$ 
the Jacobian matrices associated to the transformation $({}'c,{}'c_*) \to (c,c_*)$ following interaction rules~\eqref{eq:update1} 
and~\eqref{eq:update2}.

Now we can rewrite~\eqref{eq:strong} as a sum of gain and loss parts:
\[
    \begin{aligned}
        \partial_t f(c,t) = {}& \int_{\Omega^2}\int_{\I} 
        \Bigl(\,{}'\B_a\frac{1}{J_a}f({}'c,t)f({}'c_*,t)
                           -\B_a f({}'c,t)f({}'c_*,t) \Bigr)\, dc_*\,\eta_a\,\tilde\eta_a\\
                         {}+{}& \int_{\Omega^2}\int_{\I} \Bigl( 
                         \,{}''\B_r\frac{1}{J_r}f({}''c,t)f({}''c_*,t)
                           -\B_r f({}''c,t)f({}''c_*,t)\Bigr)\, dc_*\, \eta_r\, \tilde\eta_r,
        \end{aligned}
\]
where we indicate with $Q_a^\Sigma$ and $Q_r^\Sigma$ the operators obtained replacing 
the interaction kernels $\B_a(c,c_*)$ and $\B_r(c,c_*)$ with their approximated versions $\B_a^\Sigma(c,c_*)$ and $\B_r^\Sigma(c,c_*)$
given by
\[
\B_i^\Sigma(c, c_*) \coloneqq \min\{ \B_i(c, c_*), \Sigma \}, \quad i \in \{a, r\},
\]
where $\Sigma$ is an upper bound for $\B_i(c, c_*)$ over $\I^2$. If we now highlight the gain and loss parts 
of $Q_a^\Sigma$ and $Q_r^\Sigma$, we have
\[
\begin{aligned}
\partial_t f(c,t) &= \biggl[Q_a^{\Sigma+}(f,f) + 
f(c,t)\ev[\bigg]{\int 
\bigl(\Sigma - \B^\Sigma(c, c_*)\bigr)f(c_*,t)\, dc_*} -\Sigma 
f(c,t)\Bigr]\\
	&\hphantom{{}=} +\Bigl[Q_r^{\Sigma+}(f,f) + f(c,t)\ev[\bigg]{\int 
	\bigl(\Sigma - \B^\Sigma(c, c_*)\bigr)f(c_*,t)\, dc_*} -\Sigma 
	f(c,t)\biggr],
\end{aligned}
\]
where we define
\[
\begin{aligned}
	Q_a^{\Sigma+} &\coloneqq \ev[\bigg]{\int \B^\Sigma(c, c_*) \Bigl(\frac{1}{{}'J} f({}'c,t)\,f({}'c_*,t)\Bigr)\, dc_*},\\
	Q_r^{\Sigma+} &\coloneqq  \ev[\bigg]{ \int \B^\Sigma(c, c_*)\Bigl(\frac{1}{{}''J} f({}''c,t)\,f({}''c_*,t)\Bigr)\, dc_*}.
\end{aligned}
\]
Then, we discretize the time interval $[0, T]$ with time step $\Delta t > 0$ 
and denote as $f^n(c)$ the time approximation $f(c,n\Delta t)$ to consider 
the forward-Euler-type scheme
\[
f^{n+1} = (1 - \Sigma\Delta t) f^n  + \Sigma\Delta t \frac{\mathcal S(f^n, 
f^n)}{\Sigma},
\]
where we define
\[
\begin{aligned}
\mathcal S(f,f) &\coloneqq \biggl[Q_a^{\Sigma+}(f,f) + 
f(c,t)\ev[\bigg]{\int
\bigl(\Sigma - \B^\Sigma(c, c_*)\bigr)f(c_*,t)\, dc_*} -\Sigma 
f(c,t)\biggr]\\
	&\hphantom{{}=} +\biggl[Q_r^{\Sigma+}(f,f) + f(c,t)\ev[\bigg]{\int
	\bigl(\Sigma - \B^\Sigma(c, c_*)\bigr)f(c_*,t)\, dc_*} -\Sigma 
	f(c,t)\biggr].
\end{aligned}
\]
We remark that under the condition $\Sigma \Delta t \le 1$, $f^{n+1}$ is 
well-defined as a probability density. 

\bigskip
All the simulations performed for this test were carried out using $N = 10^5$ nodes, setting $V_a = V_r = 1$, $\beta = 0$, $\delta = \epsilon$ and $\sigma^2 = \sqrt{\delta/10}$. The initial datum is  distributed uniformly on the interval $[7, 13]$, i.e.,
\[
f(c, 0) = \frac{1}{2}\chi(c \in [7, 13]),
\]
and the scaling parameter~$\epsilon$ is set equal to $10^{-1}$, $10^{-2}$ and~$10^{-3}$, while for the parameter $\alpha$ we choose values equal to $100$, $1$ and $0.04$, where we remark that for our choice of parameters, the value of the threshold $\bar\alpha$ in estimate~\eqref{eq:uniform-alpha} is $\bar\alpha \approx 0.05000123810$.

\begin{figure}[hbpt]\centering
    \setbox0=\hbox{\includegraphics[width=0.3\textwidth]{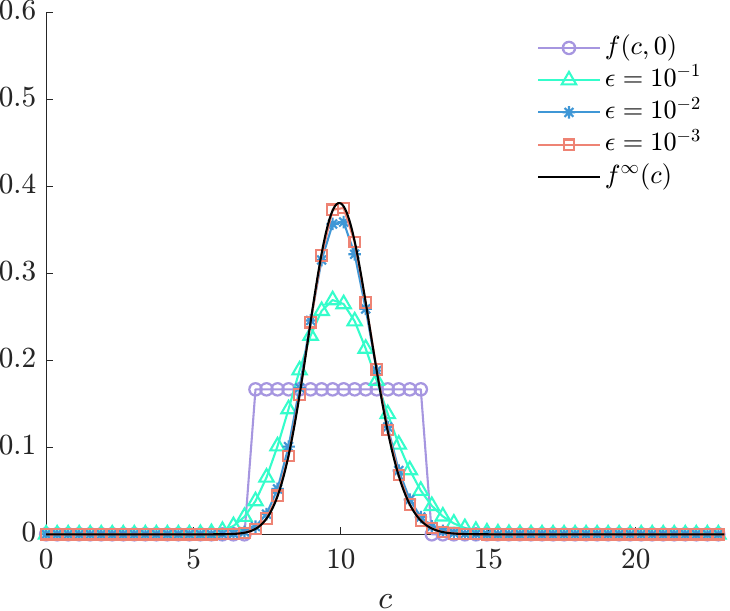}}
    \begin{tblr}{}
    \includegraphics[width=0.3\textwidth, height=\ht0]{consistency-max-100.pdf} &
    \includegraphics[width=0.3\textwidth, height=\ht0]{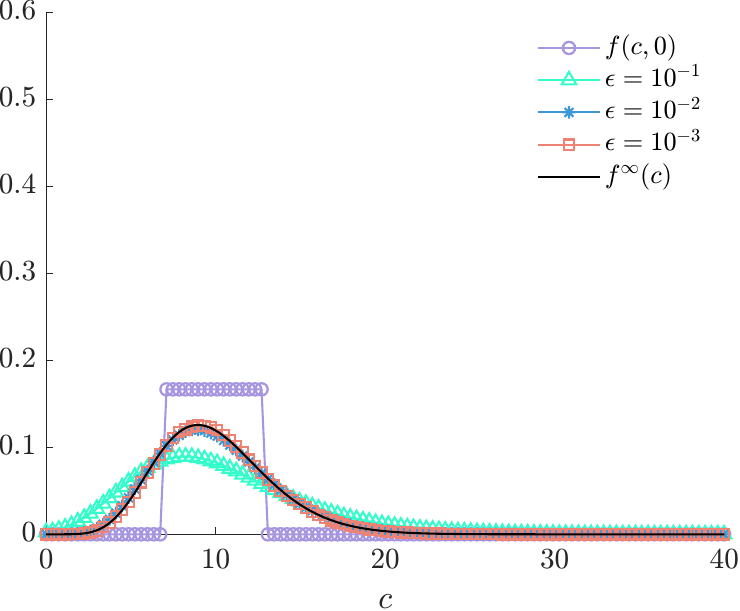} &
    \includegraphics[width=0.3\textwidth, height=\ht0]{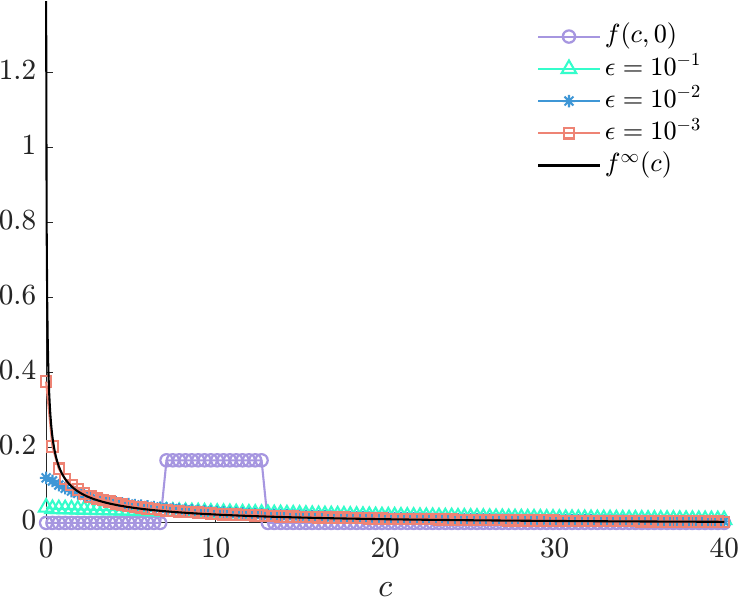} \\
    \includegraphics[width=0.3\textwidth, height=\ht0]{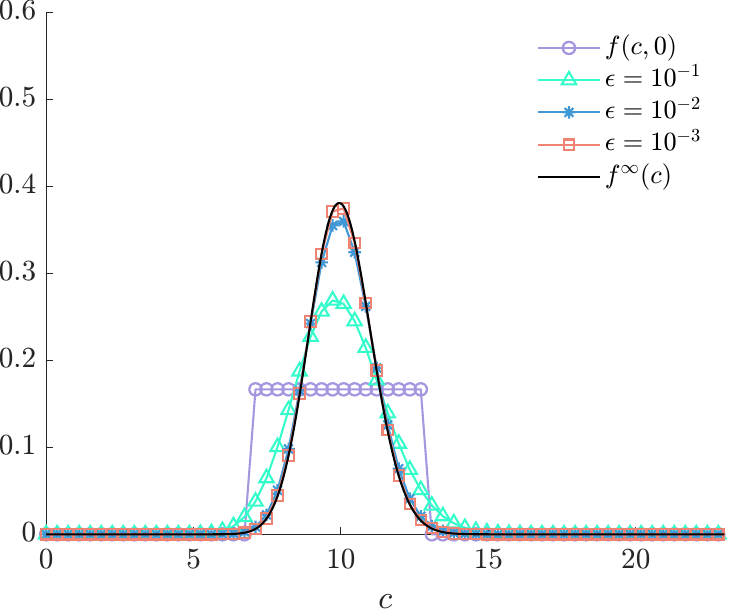} &
    \includegraphics[width=0.3\textwidth, height=\ht0]{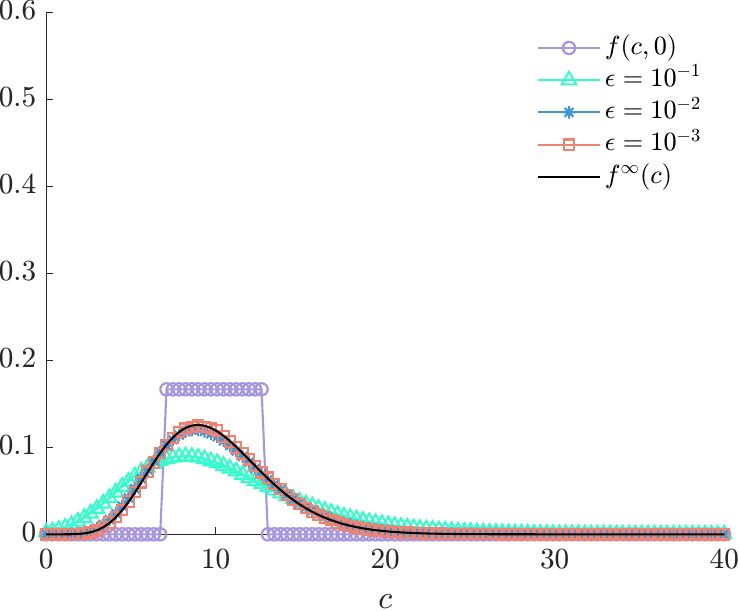} &
    \includegraphics[width=0.3\textwidth, height=\ht0]{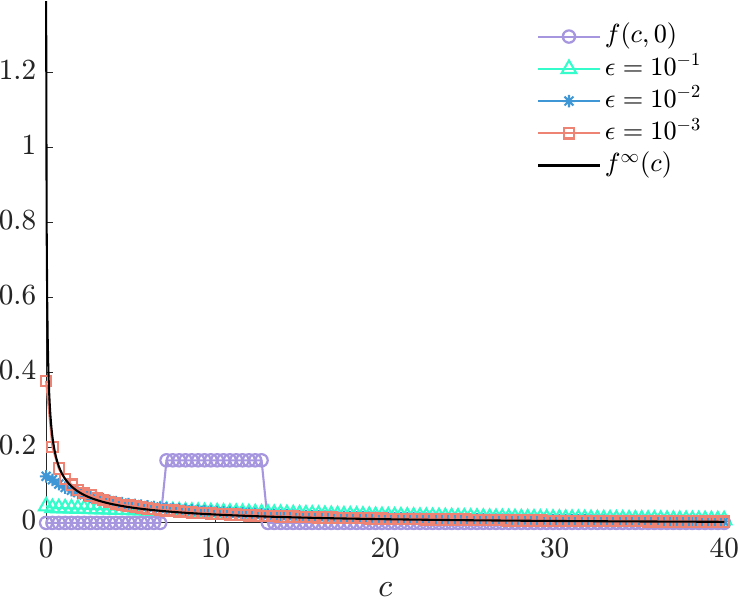}
    \end{tblr}
    \caption{Consistency of the mean-field approximation: the top
    (respectively, bottom) row shows the direct Monte Carlo simulations of 
    the Boltzmann-type model~\eqref{eq:new-wf} (respectively, 
    model~\eqref{eq:wf} for various values of $\alpha$: $100$, 
    $1$, and $0.04$, from left to right. We can see that in the limit for $\epsilon \to 0^+$ the steady state of the surrogate Fokker--Planck model is approximated accurately by the simulations of the particle models.}
    \label{fig:consistency}
    \end{figure}

In Figure~\ref{fig:consistency} we report the comparison between the numerically approximated density $f(c,T_\alpha)$ at different final times $T_\alpha$ and the analytical solution for the steady state~\eqref{eq:steady-reduced} of the surrogate model~\eqref{eq:FP}. We see that as the scaling parameter $\epsilon$ approaches zero, the large-time density approximates with ever increasing accuracy the theoretical solution, implying the consistency of the mean-field approximation of the Boltzmann-type model.

\subsection*{Test 2: rate of convergence toward equilibrium}

\begin{figure}[hbpt]
	\centering
    \setbox0=\hbox{\includegraphics[width=0.3\textwidth]{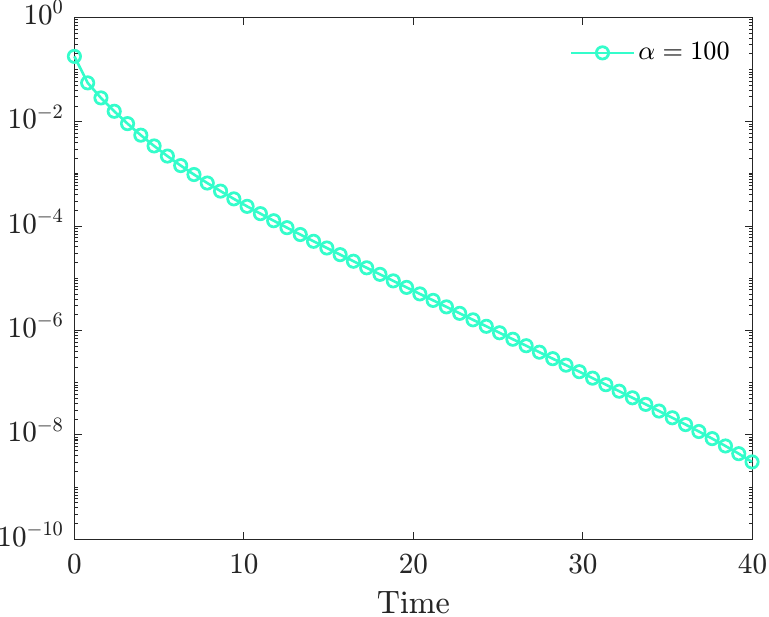}}
    \begin{tblr}{}
    \includegraphics[width=0.3\textwidth]{relative-entropy-decay-100.pdf} &
    \includegraphics[width=0.3\textwidth, height=\ht0]{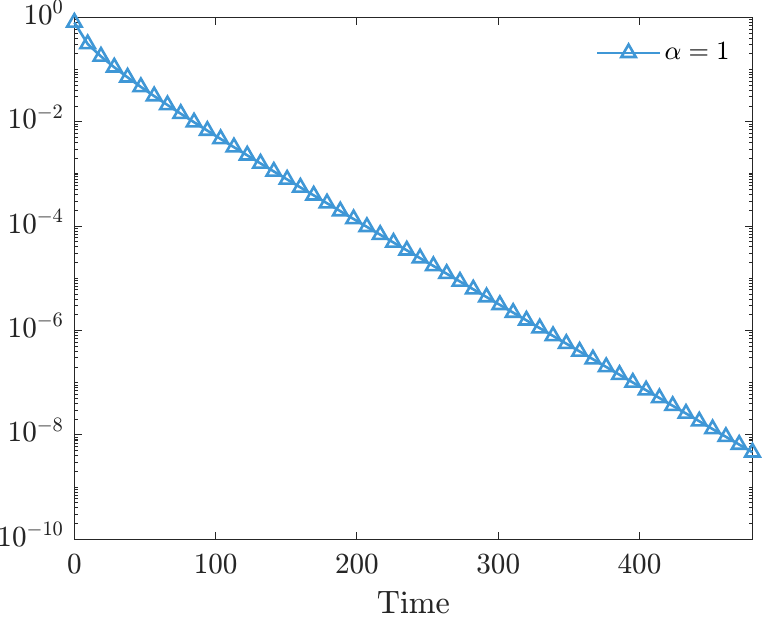} &
    \includegraphics[width=0.3\textwidth, height=\ht0]{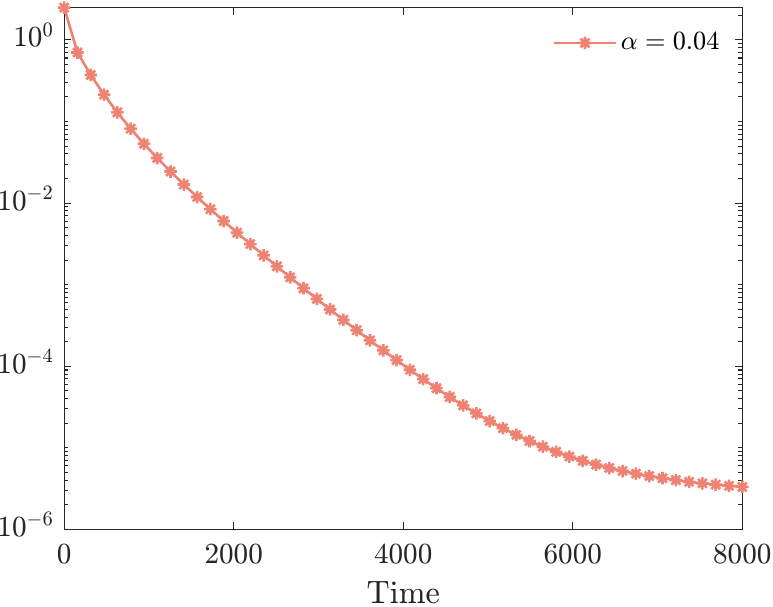}
    \end{tblr}
    \caption{Exponential convergence of the relative entropy for various values of $\alpha$: $100$, $1$ and $0.04$, from left to right in semi-logarithmic scale. We can see that a decrease in the value of $\alpha$ implies a significantly lower convergence rate, until the loss of the exponential decay.}
    \label{fig:convergence-times} 
    \end{figure}

In this test we leverage a deterministic numerical scheme of high accuracy in order to compute the approximate evolution in time of the Shannon relative entropy of the solution to the mean-field model~\eqref{eq:FP}. We refer to the same model parameters and settings as in the previous simulations. We report in Figure~\ref{fig:convergence-times} the results in semi-log scale: we see that the parameter $\alpha$ strongly influences the speed of convergence, which gets lower as $\alpha$ decreases. In particular, when $\alpha \ge \bar\alpha$ the results confirm the exponential decay of the relative entropy to zero, implying exponential relaxation toward equilibrium, while when $\alpha = 0.04 < \bar \alpha$ the decay slows down at a sub-exponential rate.

We conclude this test by describing the structure-preserving numerical scheme employed in the simulation showcased in Figure~\ref{fig:convergence-times}.

The scheme~\cite{Pareschi18} has been devised to preserve features of the solution (such as asymptotic preservation) to equations of the form
\(\label{eq:sp-general}
\left\lbrace
\begin{aligned}
\partial_t f(x,t) &= \nabla_x \cdot \bigl[\BB[f](x,t)f(x,t) + \nabla_x (D(x) f(x,t))\bigr], \qquad x \in \R^d,\\
        f(x,0) &= f_0(x).
\end{aligned}
\right.
\)
We start by rewriting equation~\eqref{eq:FP} as
\(
\partial_\tau f(c,\tau) = \partial_c \F[f](c,\tau),
\)
where $\F[\phv]$ is the Fokker--Planck flux operator
\(
\F[f](c,\tau) = \alpha\partial_c \bigl[(c - m) f(c,\tau)\bigr] + 
\frac{\lambda}{2} \partial_c^2 \bigl[((2m + \alpha)c + \alpha m)f(c,\tau)\bigr],
\)
where we applied the settings for the reduced Fokker--Planck equation~\eqref{eq:FP-reduced}. If we set $(x,c) = (c, \tau) \in \Rp \times \Rp$ and
\[
\begin{aligned}
\BB[f](c,\tau) &= \alpha\bigl[(c - m) f(c,\tau)\bigr],\\
D(c) &= \frac{\lambda}{2} \bigl[((2m + \alpha)c + \alpha m)\bigr],
\end{aligned}
\]
we see that our mean-field equation~\eqref{eq:FP-reduced} can be rewritten within the more general framework of equation~\eqref{eq:sp-general}.
Next we consider a spatially-uniform grid~$c_i \in \I$, such that $c_{i+1} - c_i = \Delta c$, and if denote
$c_{i\pm 2} = c_i \pm \Delta c/2$, we have that the discretization of~\eqref{eq:FP-reduced} can be obtained by~\cite{Pareschi18}
\(
\frac{\partial f(c,\tau)}{\partial\tau} = \frac{\F_{i+1/2}(t) - \F_{i-1/2}(t)}{\Delta c},
\)
where
\[
\begin{aligned}
\F_{i+1/2} &= \CCC_{i+1/2} \tilde f_{i+1/2} + 
                D_{i+1/2}\frac{f_{i+1}-f_i}{\Delta c},\\
\CCC_{i+1/2} &= \frac{D_{i+1/2}}{\Delta c}\int_{c_i}^{c_{i+1}}
                \frac{\B[f](c,\tau) + \partial_c D(c)}{D(c)},\\
\tilde f_{i+1/2} &= (1-\delta_{i+1/2})f_{i+1} + \delta_{i+1/2} f_i,\\
\delta_{i+1/2} &= \frac1{\lambda_{i+1/2}} + \frac1{1-\exp(\lambda_{i+1/2})},\\
\lambda_{i+1/2} &= \int_{c_i}^{c_{i+1}} \frac{\B[f](c,\tau) + \partial_c D(c)}
                    {D(c)}\, dc.
\end{aligned}
\]
Integration with respect to number of connections variable $c$ was performed with a sixth-order quadrature rule over a grid on $N_p = 801$ points on the interval $[0, 40]$, where we imposed no-flux conditions on the left and the quasi-stationary conditions on the right
\[
\frac{f_{N+1}(t)}{f_{N}(\tau)} = \exp\biggl\{\int_{c_N}^{c_{N + 1}}
                        \frac{\B[f](c,t) + \partial_c D(c)}{D(x)}\, dc\biggr\}.
\]
On the other hand, time integration was performed using an explicit 4\textsuperscript{th}-order Runge-Kutta method, with parabolic condition $\Delta \tau = (\Delta c)^2/2$ to ensure the nonnegativity of the solution.

\subsection*{Test 3: Sampling networks from the model}

In this test we show how the continuous models~\eqref{eq:wf}--\eqref{eq:new-wf} can be leveraged to obtain simple networks. The test is carried out as follows: we fix a value for the parameter~$\alpha$ and we perform a DSMC simulation of model~\eqref{eq:new-wf} with $N = 10^5$ particles until convergence to the steady state. Then, we sample from the resulting set of $N$ values a subset of $n = 5000$ values uniformly: this subset contains $n$ real values of nodes' degrees. In order to construct a simple (i.e., unweighted) network we need to transform this subset of $n$ real values in a sequence of integer numbers. To this end, we round the values stochastically under the following rule
\(\label{eq:stochastic-rounding}
\llbracket x \rrbracket \coloneqq
\begin{cases}
    \lfloor x \rfloor & \text{with probability } \lceil x \rceil - x,\\
    \lceil x \rceil & \text{with probability } x - \lfloor x \rfloor,
\end{cases}
\)
to obtain a degree sequence $S$, removing possible isolated nodes to ensure global connectedness.

The final step is to build a realization of a network with the prescribed degree sequence $S$. We use the configuration model (see e.g.,~\cite{newman03} and references therein for an exposition on the model): it gives us a realization of a multigraph with degree sequence $S$. The last step is consider the associate simple graph, obtained by removing self-loops and parallel edges.

\begin{figure}
\begin{minipage}{0.45\textwidth}
\includegraphics[width = \textwidth,
				trim=0 0cm 0 2cm]{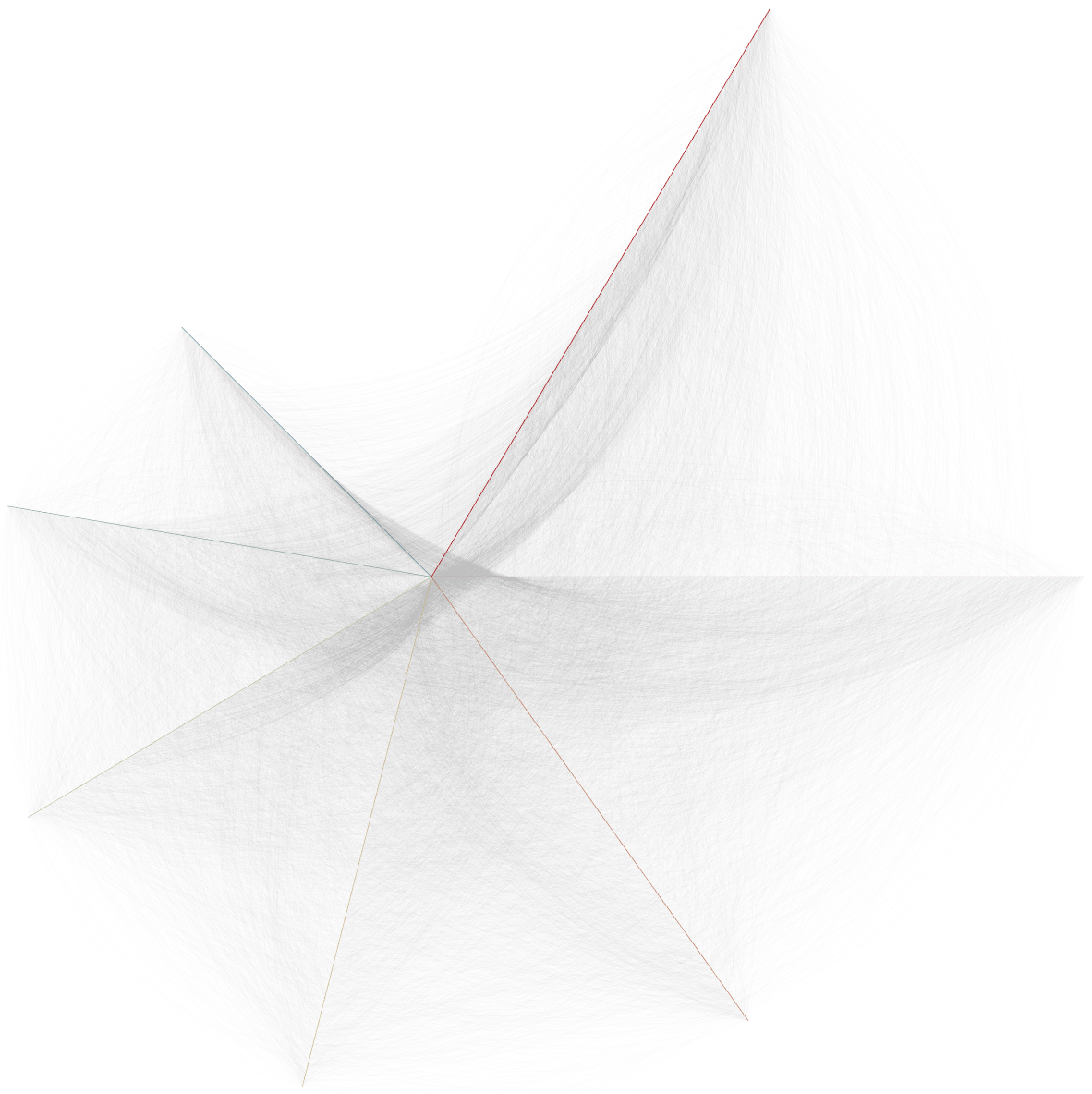}
\end{minipage}\hfill
\begin{minipage}{0.45\textwidth}
\includegraphics[width = \textwidth]{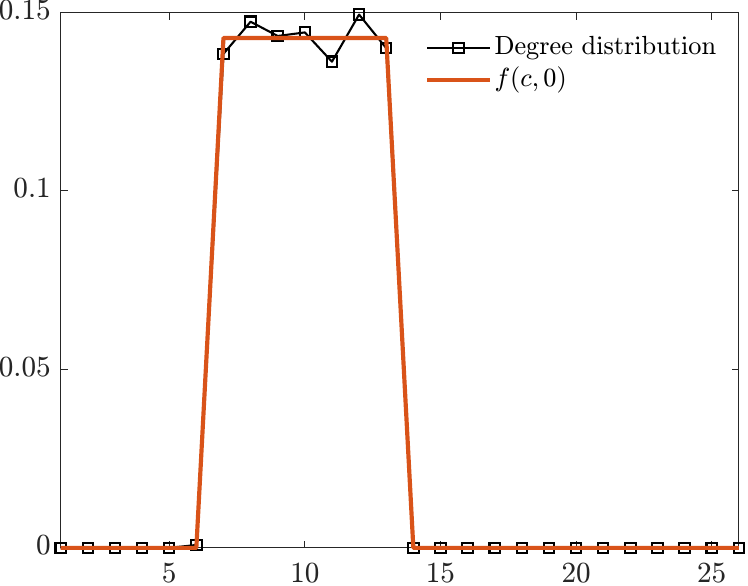}
\end{minipage}
\begin{minipage}{0.45\textwidth}
\includegraphics[width = \textwidth, trim=0 2.5cm 0 2.5cm]{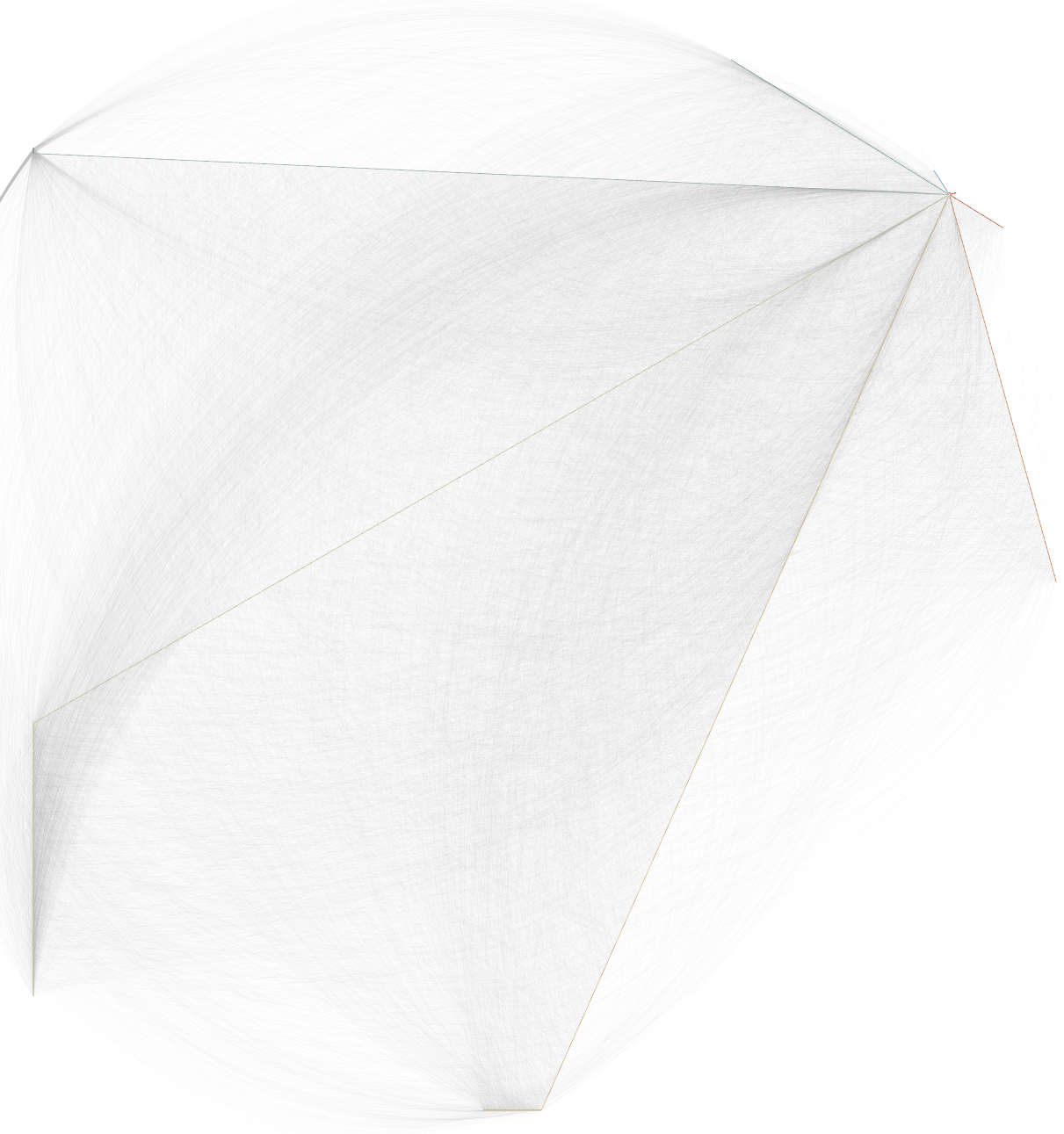}
\end{minipage}\hfill
\begin{minipage}{0.45\textwidth}
\includegraphics[width = \textwidth]{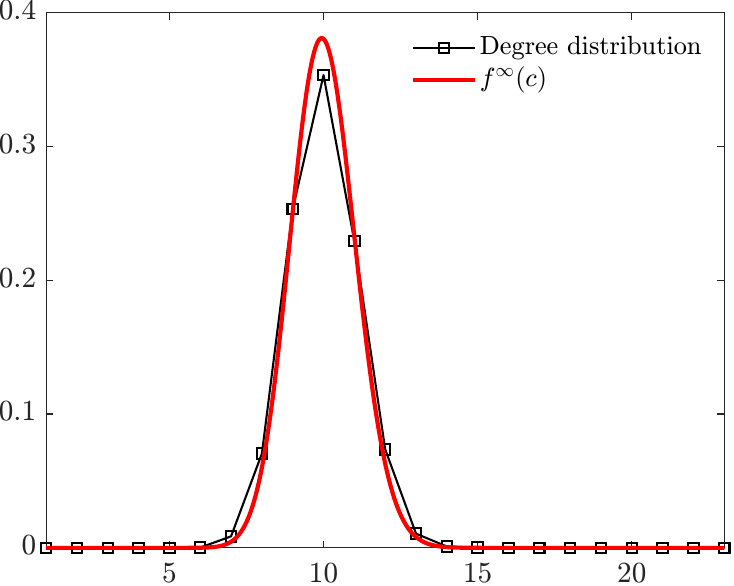}
\end{minipage}
\caption{Network realization and associated degree sequence. Top row: initial datum. Bottom row: $10^5$-nodes sample from the steady state for model~\eqref{eq:new-wf} and $\alpha = 100$. The length of each axis in the network layout on the left column is proportional to the number of nodes sharing the same degree in the network. We see accordance between the sample degree sequence and its predicted distribution.}
\label{fig:samples-1}
\end{figure}

\begin{figure}[tp]
	\centering
\begin{minipage}{0.45\textwidth}
	\includegraphics[width = \textwidth, trim=0 2cm 0 0cm]{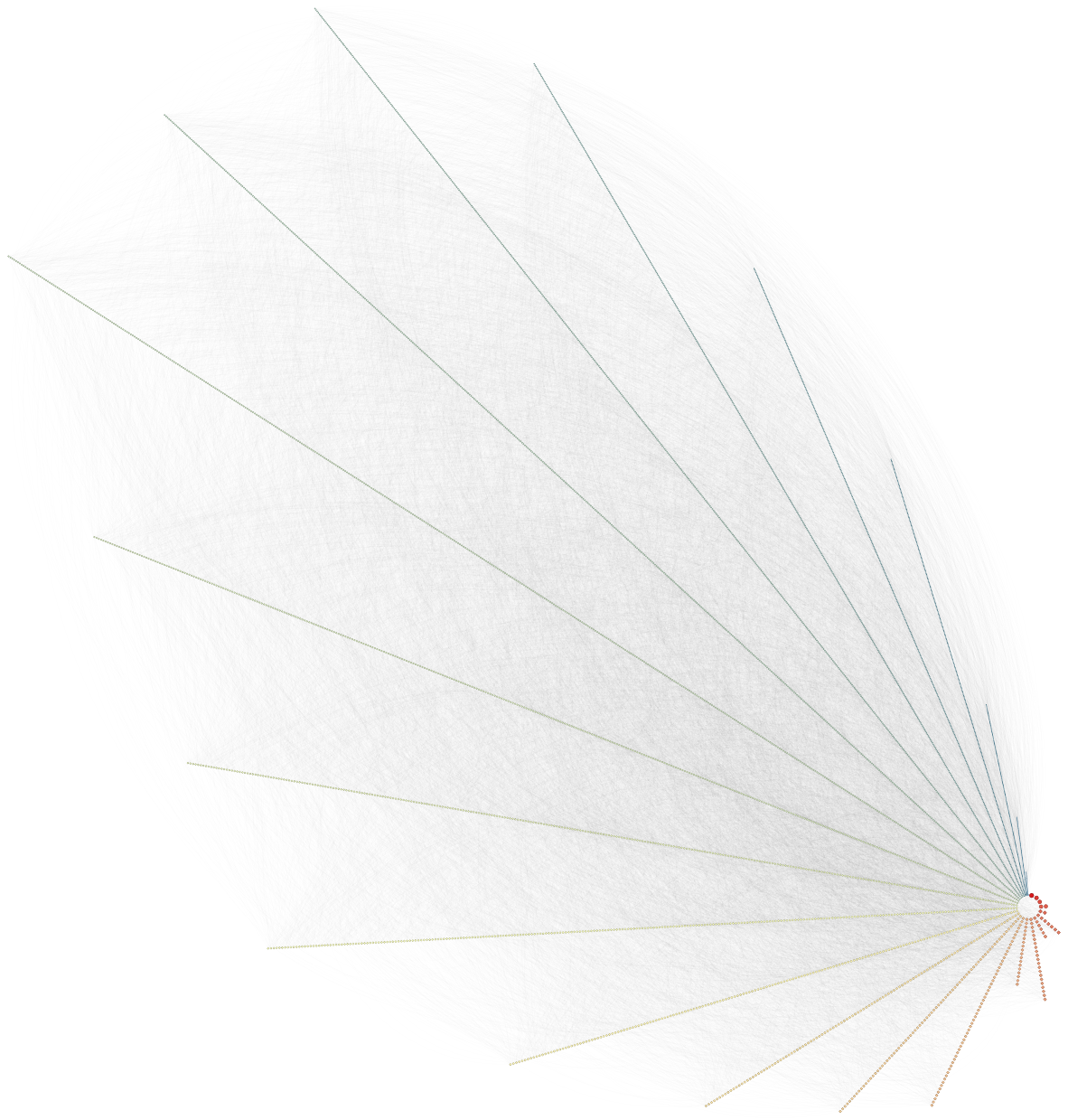}
\end{minipage}\hfill
\begin{minipage}{0.45\textwidth}
	\includegraphics[width = \textwidth]{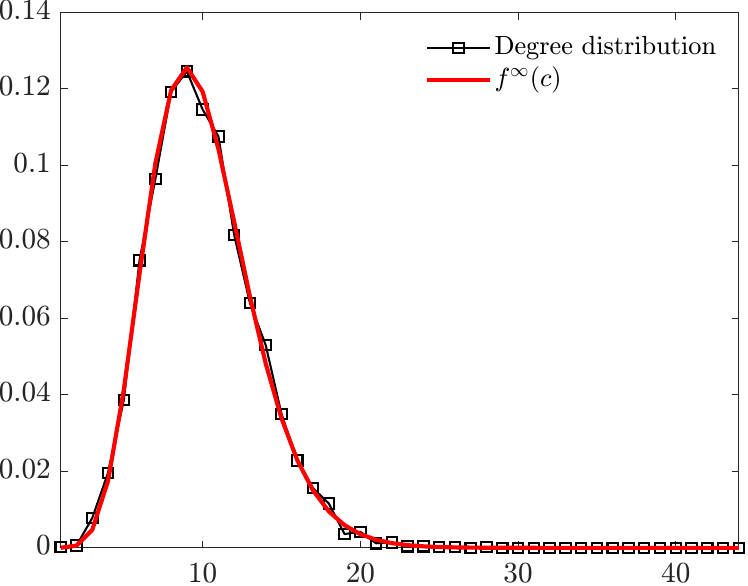}
\end{minipage}
\begin{minipage}{0.45\textwidth}
	\includegraphics[width = \textwidth, trim=0 2cm 0 0cm]{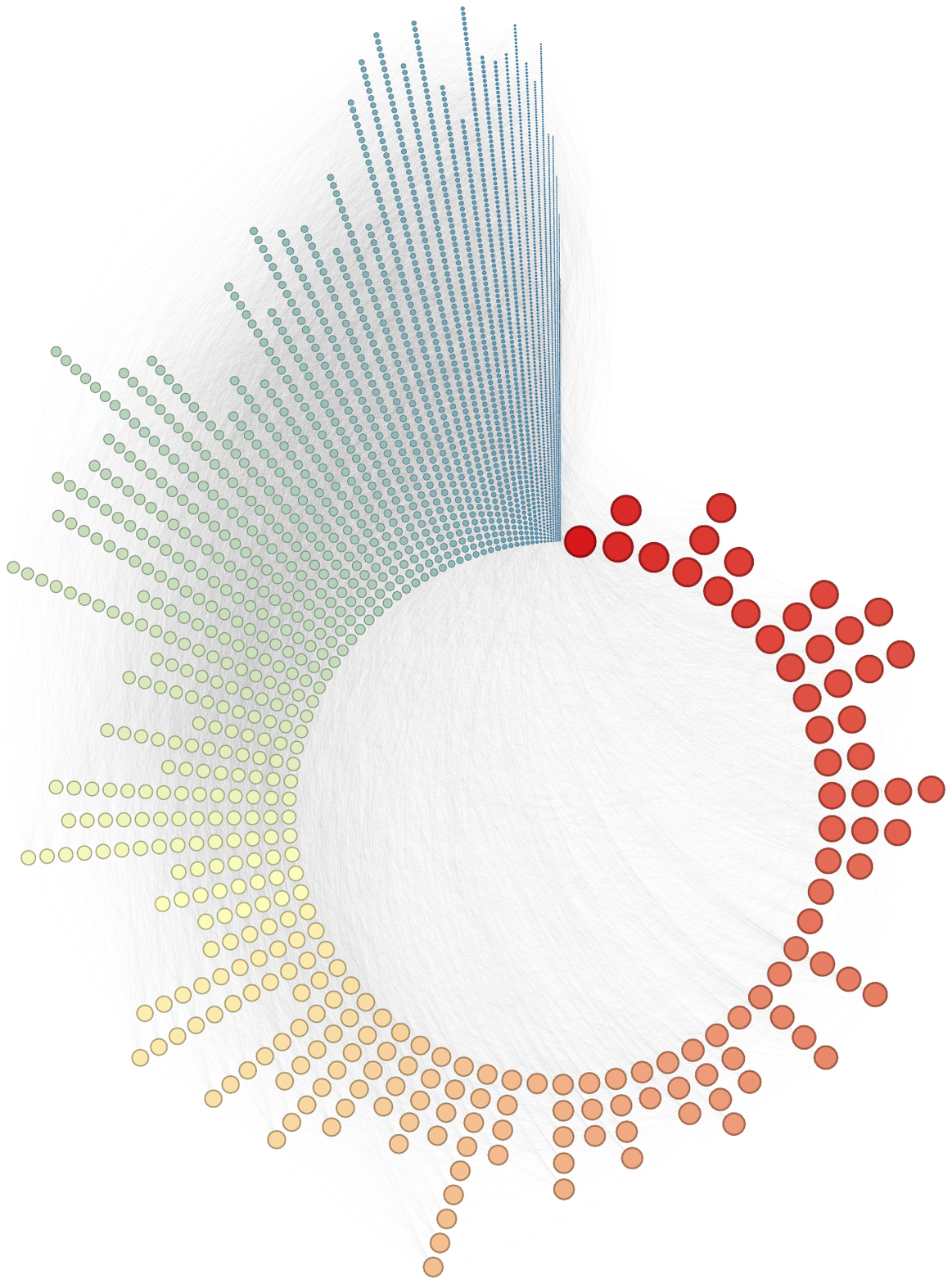}
\end{minipage}\hfill
\begin{minipage}{0.45\textwidth}
	\includegraphics[width = \textwidth]{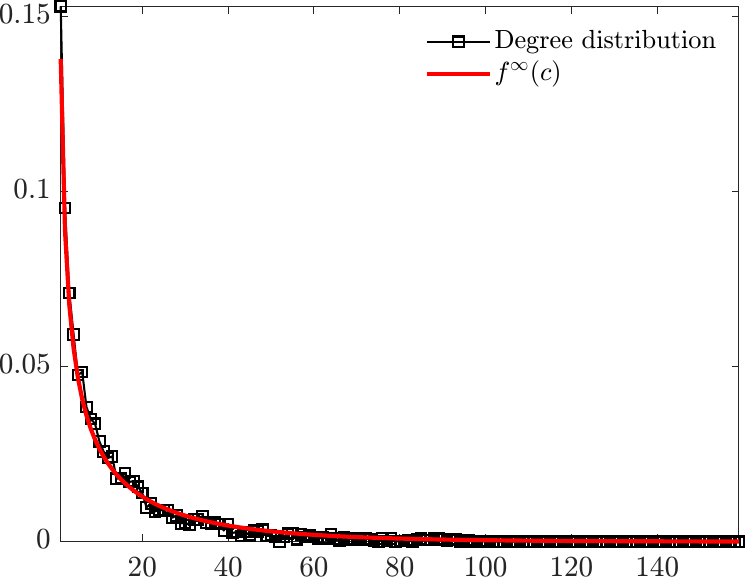}
\end{minipage}
\caption{Network realization and associated degree sequence. Top row: $10^5$-nodes sample from the steady state for model~\eqref{eq:new-wf} and $\alpha = 1$. Bottom row: $10^5$-nodes sample from the steady state for model~\eqref{eq:new-wf} and $\alpha = 0.04$. The length of each axis in the network layout on the left column is proportional to the number of nodes sharing the same degree in the network. We see accordance between the sample degree sequence and its predicted distribution. In particular, the case $\alpha = 0.04$ shows the creation of a power-law profile, with a non-negligible portion of nodes with a very high number of connections.}
\label{fig:samples-2}
\end{figure}

We use the network analysis and visualization software Gephi~\cite{GEPHI} to display the results: in particular we opt for a radial axis layout, where each node is gathered by its degree. We report the results in Figures~\ref{fig:samples-1}--\ref{fig:samples-2}.

The initial conditions and parameters are the exact same of Test 1: we report one sample for every value of $\alpha = 100$, $1$ and $0.04$, while we also display one sample for the initial datum. The radial axis layout is useful to show the degree sequence of the network graphically: both the color temperature and radius of each node in the graph are linearly proportional to its degree (the higher the degree, the greater the radius and the hotter the color temperature). Each network realization is also paired to the reconstruction of the associated degree sequence, which is itself compared to the analytical steady state: for all values of $\alpha$, the sampled network exhibit a degree sequence that accurately matches the predicted distribution $f^\infty(c)$.

\section{Conclusions}
In this work, we examine the formation of degree distributions in non-growing networks generated by a rewiring algorithm, using the framework of kinetic Boltzmann-type models. Our approach connects the microscopic dynamics of agent-based models to mean-field representations through the derivation of a Fokker--Planck-type partial differential equation (PDE). This PDE, with an affine diffusion coefficient, aligns with known master equations from discrete rewiring processes. The results offer a comprehensive framework to describe the evolution of both random and preferential attachment networks, capturing key features of scale-free behavior.

In particular, we showcased how the preferential attachment mechanism, a critical aspect of many real-world networks, can emerge naturally from a kinetic approach. By analyzing the limiting behavior of the Fokker--Planck equation, we verified that the steady-state solutions recover well-known degree distributions, such as the Poisson distribution for random graphs and the power-law distribution for scale-free networks. These results provide a robust mathematical foundation to better understand the structure of networks that do not grow in size but undergo continuous rewiring.

Furthermore, we investigated the convergence properties of the derived Fokker--Planck equation towards equilibrium. Through entropy dissipation techniques, we identified a phase transition based on the preferential attachment parameter. This transition delineates different regimes for the rate of convergence, either exponential or algebraic, depending on network characteristics. This finding enriches the understanding of the long-term behavior of non-growing networks and offers insights into how the underlying dynamics influence their stability and structure.

Future research could explore more generalized settings, such as dynamic or time-varying networks, temporal connectivity patterns, and structured graph functionals, and examine the impact of more intricate rewiring algorithms. Additionally, the framework presented here opens up possibilities for further theoretical developments in the study of non-constant diffusion processes and their applications to large-scale network phenomena.

\section*{Acknowledgments}
This work has been written within the activities of the GNCS and GNFM groups of INdAM. J.F. and M.Z. acknowledge the support of the Italian Ministry of University and Research (MUR) through the PRIN-2022PNRR project (No.\ P2022Z7ZAJ). The research of L.P. has been supported by the Royal Society under the Wolfson Fellowship ``Uncertainty quantification, data-driven simulations and learning of multiscale complex systems governed by PDEs'' and by the Italian Ministry of University and Research (MUR) through the PRIN-2022 project (No.\ 2022KKJP4X). The partial support by ICSC -- Centro Nazionale di Ricerca in High Performance Computing, Big Data and Quantum Computing, funded by European Union -- NextGenerationEU is also acknowledged by M.Z. and L.P.

\printbibliography
\end{document}